\documentclass[ aip, amsmath,amssymb, reprint,floatfix]{revtex4-1}

\pdfoutput=1

\usepackage{multirow}
\usepackage{graphicx}
\usepackage{dcolumn}
\usepackage{bm}

\usepackage[utf8]{inputenc}
\usepackage[T1]{fontenc}
\usepackage{mathptmx}
\usepackage{etoolbox}
\usepackage{hyperref}

\makeatletter
\def\@email#1#2{%
 \endgroup
 \patchcmd{\titleblock@produce}
  {\frontmatter@RRAPformat}
  {\frontmatter@RRAPformat{\produce@RRAP{*#1\href{mailto:#2}{#2}}}\frontmatter@RRAPformat}
  {}{}
}%
\makeatother

\begin{document}

\preprint{AIP/123-QED}

\title[Phase behavior of metastable water from large-scale simulations of a quantitative accurate model: The liquid-liquid critical point.]{Phase behavior of metastable water from large-scale simulations of a quantitative accurate model: The liquid-liquid critical point.}

\author{Luis Enrique Coronas}
\altaffiliation[Current address: ]{Laboratoire de Biochimie Théorique, UPR 9080 CNRS, 13 rue Pierre et Marie Curie, 75005 Paris, France}
 
\author{Giancarlo Franzese}

\affiliation{Secci\'o de F\'isica Estad\'istica i Interdisciplin\`aria, Departament de F\'isica de la Mat\`eria Condensada, Facutat de F\'isica, University of Barcelona, Mart\'i i Franqu\`es 1, Barcelona 08028, Spain}
\affiliation{Institute of Nanoscience and Nanotechnology (IN2UB), University of Barcelona, Mart\'i i Franqu\`es 1, Barcelona 08028, Spain}
\altaffiliation[On leave: ]{Max Planck Institute for the Physics of Complex Systems, N\"othnitzer Straße 38, Dresden, 01187, Germany}

\email{gfranzese@ub.edu}

\date{\today}

\pacs{}

\begin{abstract} 
Water's unique anomalies are vital in various applications and biological processes, yet the molecular mechanisms behind these anomalies remain debated, particularly in the metastable liquid phase under supercooling and stretching conditions. Experimental challenges in these conditions have led to simulations suggesting a liquid-liquid phase transition between low-density and high-density water phases, culminating in a liquid-liquid critical point (LLCP). However, these simulations are limited by computational expense, small system sizes, and reliability of water models.
Using the FS model, we improve accuracy in predicting water's density and response functions across a broad range of temperatures and pressures. The FS model avoid {\it by design} first-order phase transitions towards crystalline phases, allowing thorough exploration of the metastable phase diagram. We employ advanced numerical techniques to bypass dynamical slowing down and perform finite-size scaling on systems significantly larger than those used in previous analyses.
Our study extrapolates thermodynamic behavior in the infinite-system limit, accurately demonstrating the existence of the LLCP in the 3D Ising universality class at $T_C = 186\pm4$ K and $P_C = 174\pm14$ MPa, following a liquid-liquid phase separation below 200 MPa. These predictions align with recent experimental data and more sophisticated models, highlighting that hydrogen bond cooperativity governs the LLCP and the origin of water anomalies.
Moreover, we observe that the hydrogen bond network exhibits substantial cooperative fluctuations at scales larger than 10 nm, even at temperatures relevant to biopreservation. These findings have significant implications for fields such as nanotechnology and biophysics, offering new insights into water's behavior under varied conditions.
\end{abstract}

\maketitle

\section{\label{sec:Introduction}Introduction}

Water is essential in our lives, but its complex nature still raises many unresolved questions \cite{RevModPhys.88.011002, Handle2017, Palmer:2018ac, Gallo:2021wx}. It is unique compared to other liquids for its more than 60 anomalies  \cite{Chaplin}, such as a density maximum at around 4°C, or the large increase of its response functions upon cooling, when instead a decrease would be expected for usual liquids
\cite{debenedetti-book, RevModPhys.88.011002, Gallo:2021wx}.

Several scenarios have been suggested to explain these properties \cite{Speedy82, llcp, Sastry:1996aa, Angell2008, Gallo:2021wx}. The theory \cite{Stokely2010} has shown that all these scenarios stem from a general thermodynamic description and differ in the ratio between the intensity of the cooperative component \cite{Barnes1979} and the covalent component \cite{Shi:2018ac} of the hydrogen bond (HB) network. 
The estimate of this ratio supports \cite{Stokely2010} that water follows the scenario originally proposed by Poole et al. \cite{llcp} based on molecular dynamic simulations. Poole and coworkers conceived that water undergoes a liquid-liquid phase transition (LLPT) between low-density liquid (LDL) and high-density liquid (HDL) phases, ending in a liquid-liquid critical point (LLCP) in the supercooled region. Several water models, ranging from atomistic to machine-learned {\it ab initio} quality force fields \cite{Dhabal_Kumar_Molinero_2024}, have supported this hypothesis over the years, generating a large amount of numerical data that have dissipated criticisms and doubts about the theoretical possibility of such a scenario \cite{Palmer:2014uq}.
Yet, no experiment has shown the existence of a liquid-liquid phase transition in deeply supercooled water due to the difficulty in avoiding crystallization \cite{Kim978}.

To date, substantial evidence supports the LLCP hypothesis based on multiple experiments and computational approaches. On the experimental side, several techniques have been used to measure structural changes that reflect the  LLPT. For example, neutron scattering has been applied to the supercooled liquid of a Zr-Cu-Al-Ag alloy \cite{Dong2021}, while Raman spectroscopy has been used to study ionic liquids \cite{Harris2021} or aqueous solutions \cite{C9CP06082K, Suzukie2113411119}. However, the results obtained from experiments on aqueous solutions, which prevent rapid crystallization, cannot be straightforwardly extrapolated to water \cite{BachlerPRE2020}. On amorphous ice, X-ray experiments have shown results consistent with a first-order phase transition \cite{Winkel2011, Perakis:2017aa}. More recently, Kim et al. conducted experiments on micro-sized liquid water droplets \cite{Kim2017, Caupin:2018aa, Kim:2018aa} and bulk \cite{Kim978}. Their results show a sudden change in the structure factor at one order of magnitude shorter times than subsequent crystallization, which is consistent with LLPT for temperatures around $(205\pm10)$ K and pressures between 1 atm and 350 MPa \cite{Kim978}, and the presence of an LLCP at positive pressure \cite{Kim2017, Kim978, Nilsson:2022tx}.

On the computational side, different models support the LLCP in water, either rigid or flexible. The rigid ST2 \cite{llcp}, has been rigorously proven to have an LLPT \cite{Liu2009, Sciortino2011, Liu2012, Kesselring2012, Kesselring2013, Palmer:2014uq}, as well as TIP4P/2005 \cite{Debenedetti289, Sciortino2024}, TIP4P/Ice \cite{Debenedetti289, Sciortino2024, 10.1063/5.0049299, Espinosa2023}, 
TIP4P \cite{corradini:134508},
and patchy models  \cite{Buldyrev:2015uq, Neophytou:2022wc}.
Flexible models, such as the polarizable WAIL~\cite{Weis:2022ug} and the iAMOEBA \cite{Wang_amoeba_2013}, which employs point multipole electrostatics and an approximate description of electronic polarizability, with up to three-body terms \cite{10.1063/1.4963913}, also have an LLCP.

Other models that display LLCP and include many-body interactions to describe the cooperative nature of hydrogen bonding are the coarse-grained water monolayer with up to five-body interactions \cite{Bianco2014}, the EB3B with up to three-body interactions \cite{Hestand:2018aa, Ni2016} using the TIP4P/2005 \cite{Abascal:2005bh} as two-body reference,  the coarse-grained machine-learned ML-BOP \cite{Dhabal_Kumar_Molinero_2024, 10.1063/5.0197613}, with up to four-body interactions \cite{Chan2019}, and, just released, the DNN@MB-pol model \cite{Sciortino_Zhai_Bore_Paesani_2024} that is a machine-learning surrogate of the MB-pol, a model with mean-field-like representations of many-body electrostatic interactions integrated with machine-learned representations of short-range quantum-mechanical effects \cite{BorePaesani2023}.

Furthermore, the phenomenological two-states equation of state (TSEOS) fits remarkably well to several water models \cite{10.1063/1.4973546, Gartner2020, Eltareb:2022wk, Weis:2022ug} by assuming the LLCP hypothesis \cite{Singh2016}. Although the TSEOS fitting method does not constitute definite proof, it is beneficial if simulation data is not available at the hypothesized critical point, as it shows that the model is at least consistent with the presence of an LLCP.

Simple models can help us understand the physical mechanisms that lead to the LLCP. For example, the role of HB cooperativity, or spatial correlation, has been emphasized both in continuous \cite{StanleyTeixeira1980, PhysRevE.49.2841} and two-states model  \cite{PhysRev.139.A758, 10.1063/1.1841150, PhysRevE.62.6968, Franzese:2000aa, FSPhysA2002, Stokely2010}. Based on available experimental data, cooperative models \cite{Stokely2010, Tanaka:2020aa} have predicted the LLCP at positive pressure.

On the other hand, commonly used models of water, such as SPC/E  or mW, do not display the LLCP, showing no necessary relation with the water anomalies \cite{Gallo:1996dq, Limmer2011}. In other models, e.g., the TIP4P/2005, severe finite-size effects are an issue \cite{Overduin:2013fk}, and extremely long simulations of more than 50 $\mu$s for 1000 molecules are necessary to obtain well-converged density distributions for deeply supercooled water \cite{Debenedetti289}. Despite their differences in the supercooled water, these models qualitatively reproduce the experiments around ambient conditions. Hence, the challenge to understanding the metastable water phase diagram is to explore a model that a) can {\it quantitatively} reproduce water thermodynamics over a broad range of temperatures and pressures, b) for system sizes large enough to perform the necessary finite-size scaling analysis, and c) at an accessible computational cost. 

In this study, we examine the phase diagram of metastable liquid water for the FS model, which has exceptional accuracy in replicating the experimental data of density and response functions of water in a broad region of thermodynamic conditions, expanding up to 60 degrees around ambient conditions and 40 degrees up to 50 MPa \cite{Coronas-2024}. 

Although the model can be defined in such a way to include crystal phases and polymorphism \cite{Vilanova2011, Bianco:2014fk}, here we consider the formulation that {\it by design} has no first-order phase transition toward the crystalline phases. This embedded feature is an advantage in the study of metastable liquid water because it
allows one to sample the free energy landscape in a highly efficient way.

Additionally, the model's simple definition enables the use of advanced numerical techniques, avoiding dynamical slowing down near critical points or approaching a glassy state, which is typically found in supercooled water. Consequently, the FS model emerges as an intriguing candidate to explore the phase diagram of metastable water in regions currently inaccessible to experiments.

Another compelling advantage of the FS model is that it allows for efficient equilibration of free energy calculations for systems up to 10 million molecules using common GPUs \cite{CoronasThesis}.
All these features are crucial for analyzing metastable water because small-size effects, freezing in a glassy state, and crystallization typically hamper similar studies.

Using a state-of-the-art procedure \cite{Liu2010, Bianco2014, Debenedetti289, Weis:2022ug}, we equilibrate the metastable water under extreme conditions and systematically study finite but large systems. This allows us to extrapolate the thermodynamic behavior in the infinite-system limit, following the finite-size scaling theory, and demonstrate, with great accuracy, the existence of the LLCP for the model in the thermodynamic limit. We show that the LLCP belongs to the 3D Ising universality class, corroborating the thermodynamic consistency of the result. We compare our prediction with other models, finding similarities that suggest that the hydrogen bond network can develop significant cooperative fluctuations at scales relevant to nanotechnology applications and biological systems. 

\section{\label{sec:ModelAndMethod}Monte Carlo method}

\subsection{\label{sec:simulationsPhaseDiagram}Large-scale Monte Carlo free energy calculations for metastable bulk water}

We perform Monte Carlo (MC) calculations of the free energy of the FS water model in 3D with parameters able to reproduce quantitatively the equation of state of liquid water over a temperature range of approximately 60 degrees around ambient conditions at atmospheric pressure and about 40 degrees up to 50 MPa \cite{Coronas-2024}.
The FS model is defined in Appendix~\ref{app:TheFSModel}.
We consider a constant number of molecules $N$, pressure $P$, and temperature $T$ in a variable cubic volume $V_{\rm Tot}$ with periodic boundary conditions.

We calculate the equation of state along isobars in the range of $-540$ MPa to 260 MPa, separated by intervals $\Delta P \leq 50 {\rm ~MPa}$. The range of temperatures is 187 K to 470 K, with simulated thermodynamic points at variable resolution, $0.014 {\rm ~K} \leq \Delta T \leq 14 {\rm ~K}$, depending on the region of interest. 
The data presented here include results for systems of up to $N=$32,768 water molecules. We tested on several selected state points (not shown) that thermodynamic quantities calculated for $N =$262,144 are statistically similar to those for $N=$32,768 within our numerical precision but require roughly ten times longer to compute.

We apply the sequential annealing protocol along isobars, starting at high $T$ and lowering the temperature in stages, allowing the system to equilibrate at each stage. We check the equilibration by verifying the validity of the fluctuation-dissipation theorem for the response functions defined in the following.
We collect data by averaging over $10^4$ - $10^5$ MC steps after equilibration, corresponding to $10^3$ - $10^4$ independent configurations depending on the state point. 

For $N=$32,768, we apply the Metropolis algorithm at temperatures $T \geq 215$ K. At lower temperatures, we use the Swendsen-Wang cluster 
algorithm \cite{Mazza2009} based on the site-bond correlated percolation \cite{Bianco:2019aa}. 
This approach allows us to overcome the slowing down due to the approaching of the glassy state \cite{FdlSJPCM2009} that characterizes water dynamics at these values of $T$ and $P$ \cite{debenedetti_stanley, Sciortino2011, Kesselring2012}.
We optimize the performance of our simulations by implementing parallelization in GPUs using CUDA~\cite{CoronasThesis}.

\subsection{\label{sec:MCsimulationsCriticalRegion}Monte Carlo calculations in the critical region}

As described in the section \ref{sec:Results}, we estimate the loci in the $(T, P)$ plane where the response functions of metastable water have maxima and verify if, in the region where they converge, the fluctuations exhibit the expected finite-size behavior near a critical point. To achieve this, we compare calculations for systems with $N$ up to 4,096 water molecules. These sizes were selected to ensure that the coexisting HDL and LDL states could transition frequently within reasonable simulation time. Larger systems exhibit single transitions from HDL to LDL, consistent with the expected exponential decrease of the transition rate as the free energy barrier between the phases increases.

We apply the sequential annealing protocol starting at $T=270$~K at each $ P$ and $ N$, using the Wolff cluster algorithm \cite{Wolff, Mazza2009}.
We adapt the temperature resolution to the $T$-derivative of the calculated quantity, ranging from $\Delta T=0.014$~K to 14~K. 
We observe that the Wolff algorithm's implementation on CPUs outperforms Swendsen-Wang, particularly for small $N$. 
After a preliminary isobaric scan, we select the $T$ for each $N$ with the most frequent transitions between the HDL and LDL states and adjust the simulation time. We find that the time necessary to observe multiple HDL-LDL transitions rapidly increases with $N$, ranging from $10^5$ to $10^8$ MC steps for $N$ going from  512 to 4,096 (see Table \ref{table:simulation_times}). For $N=$4,096, we observe coexistence only for $P \geq$ 110~MPa.

\subsection{\label{sec:estimationLLCP} LLCP and universality class analysis}

To rigorously prove the presence of an LLCP, we need to identify the correct order parameter (o. p.) that describes the phase transition. In fluid-fluid phase transitions, the critical point belongs to the 3D Ising universality class with a {\it mixed-field} o. p. that combines the number density with the energy density to account for the lack of symmetry in the critical density distribution \cite{Wilding1996, Liu2010}. Therefore, 
a definitive proof that the FS model displays an LLCP is that the fluctuations of the correct o. p. behave in a manner consistent with those of the magnetization of the 3D Ising model at its critical point, as shown already for soft-core isotropic anomalous fluids \cite{Gallo:2012fk}, rigid~\cite{Kesselring2012, Debenedetti289}, and flexible~\cite{Weis:2022ug} water models, as well as the Stillinger-Weber model for silicon~\cite{Goswami2022}. The same approach has also been used to show that the confined FS monolayer has an LLCP belonging to the 2D Ising universality class \cite{Bianco2014}.

In the $NPT$-ensemble, based on finite-size scaling theory for density-driven fluid-fluid phase transitions \cite{Wilding1996}, the o. p. is 
$M(s)\equiv \bar{\rho} + s \bar{e}$, where $\bar{\rho}$ and $\bar{e}$ are dimensionless density and energy, respectively, and $s$ is the so-called {\it mixing parameter}. 
This linear combination symmetrizes the probability distribution of the coexisting states that for water at the LLCP would correspond to HDL and LDL phases  \cite{Bianco2014, Bianco:2019aa}.

We use the histogram reweighting method~\cite{Panagiotopoulos2000, Bianco2014} to find the critical temperature $T_C$, critical pressure $P_C$, and $s$ values that allow us to fit the fluctuations of $M$ with those expected for the Ising 3D critical point. 
To do this, we calculate by MC the histogram $H_i(T_i, P_i;e,\rho)$ of visited configurations with given molar energy $e$ and density $\rho$ for state points ($T_i$, $P_i$) within the critical region.
By normalizing $H_i$ we provide a finite-size (numerical) approximation of the probability density distribution $Q(e, \rho)$ of states with given $(e, \rho)$ for the system in the thermodynamic limit.
We then combine a set of $H_i$ by the histogram reweighting method to calculate $H(T,P;e,\rho)$ for any $(T, P)$ near ($T_i$, $P_i$) \cite{Bianco2014}. 

We optimize $T_C$, $P_C$, and $s$ based on $H(T, P;e,\rho)$ in the critical region to ensure that the integral of $H(T_C, P_C;e,\rho)$ over $(e,\rho)$ corresponding to the same $M(s)$––i.e., $H(T_C, P_C; M(s))$––has a bimodal distribution centered around a value that we indicate as $M_C(s)$. Then, as discussed in Appendix~\ref{app:OrderParameter}, we determine a rescaled version of $M(s)$, $x=x(s)$ with zero mean and unit variance, which probability density distribution $Q(x)$ at $\{T_C, P_C\}$ approximates that of the 3D Ising model o. p. $m_I$ at criticality, $Q_3(m_I)$. We iterate the process and selection of $T_C$, $P_C$, and $s$ to minimize the difference between $Q$ and $Q_3$.

\section{\label{sec:Results}Results}

\subsection{\label{sec:ResultsBulkFS}Density, enthalpy, and HB network.} 

\begin{figure}
    \centering
    \includegraphics[scale=0.24]{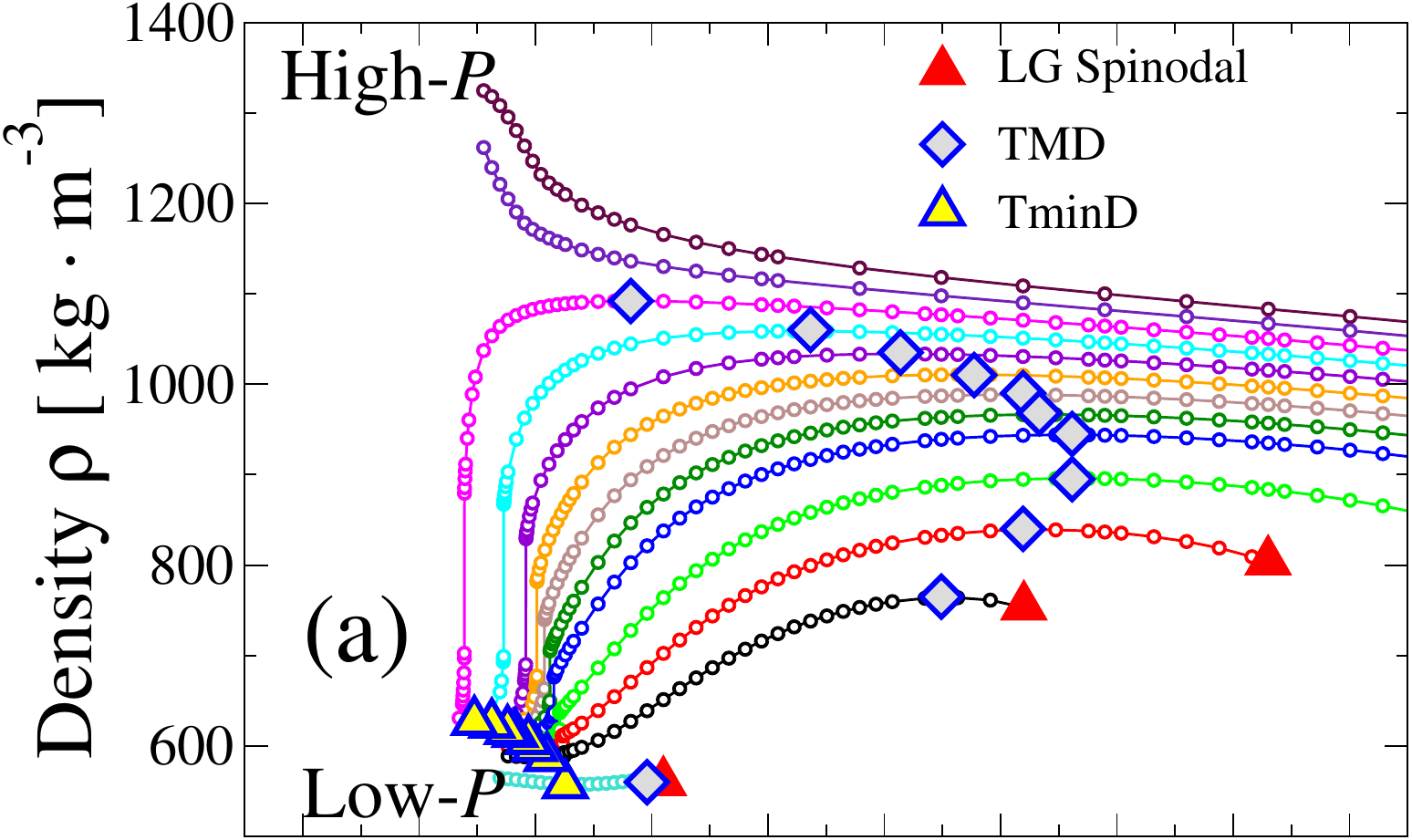} \ \ \ 
    \includegraphics[scale=0.24]{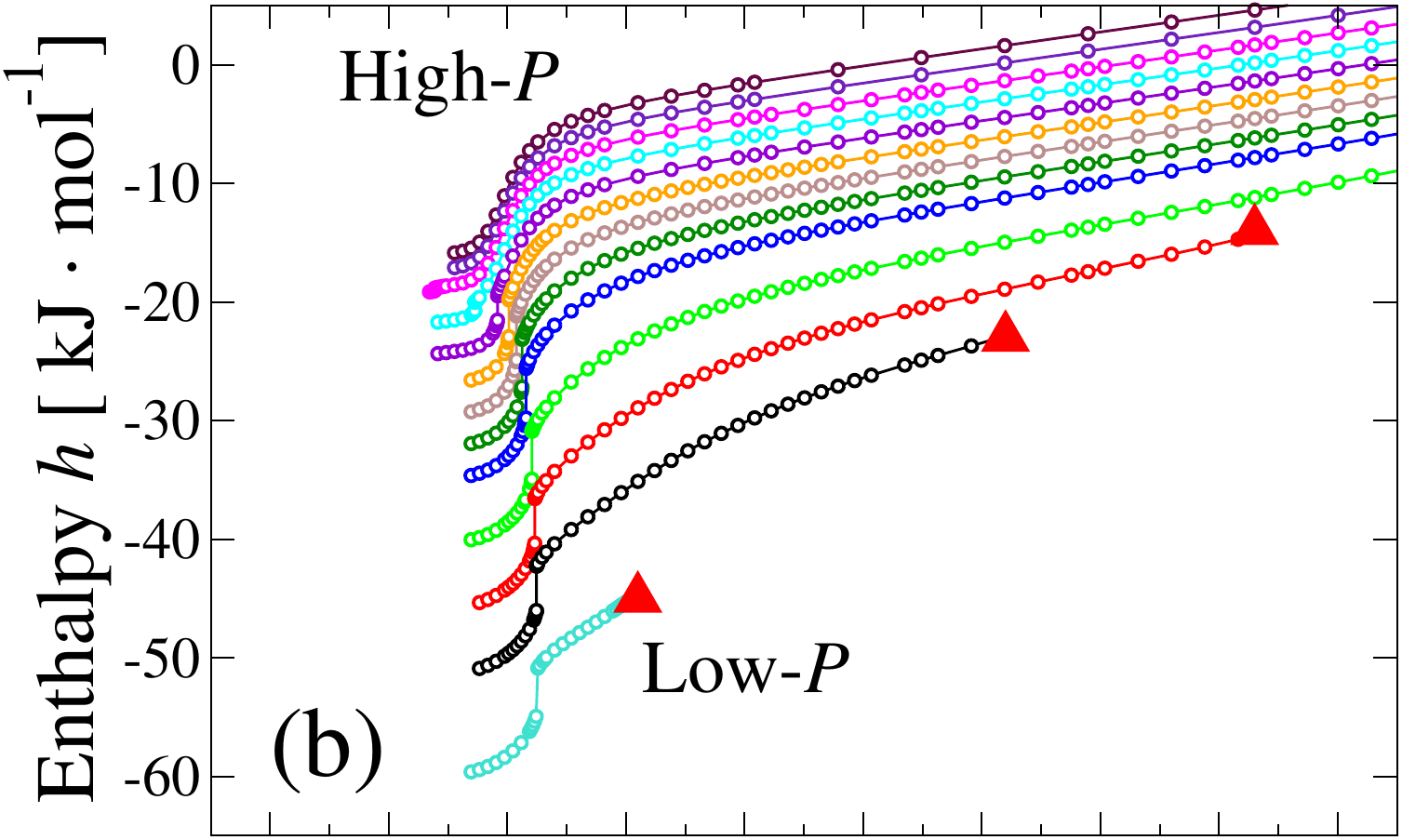}
    \includegraphics[scale=0.24]{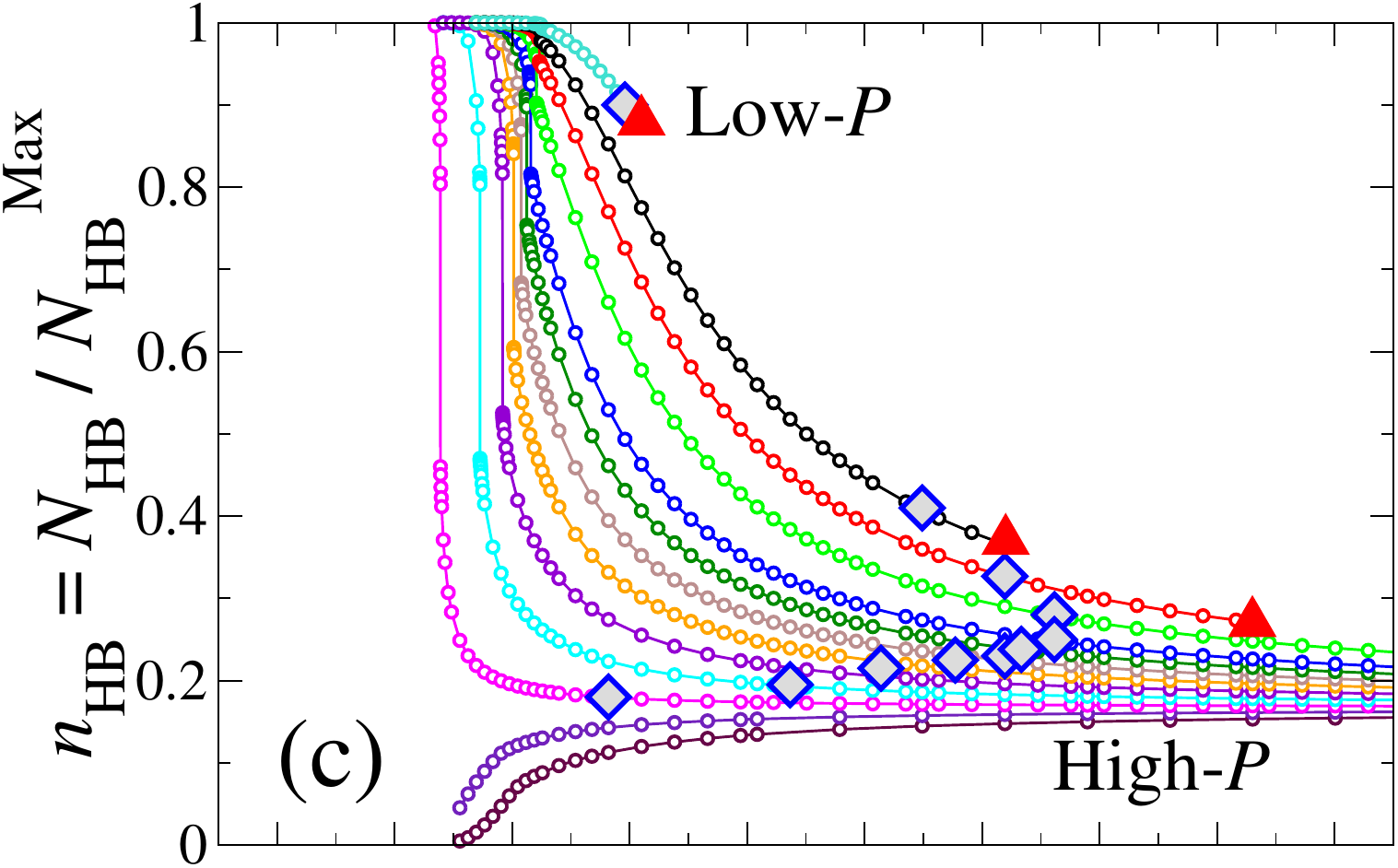} \  \ \ 
    \includegraphics[scale=0.24]{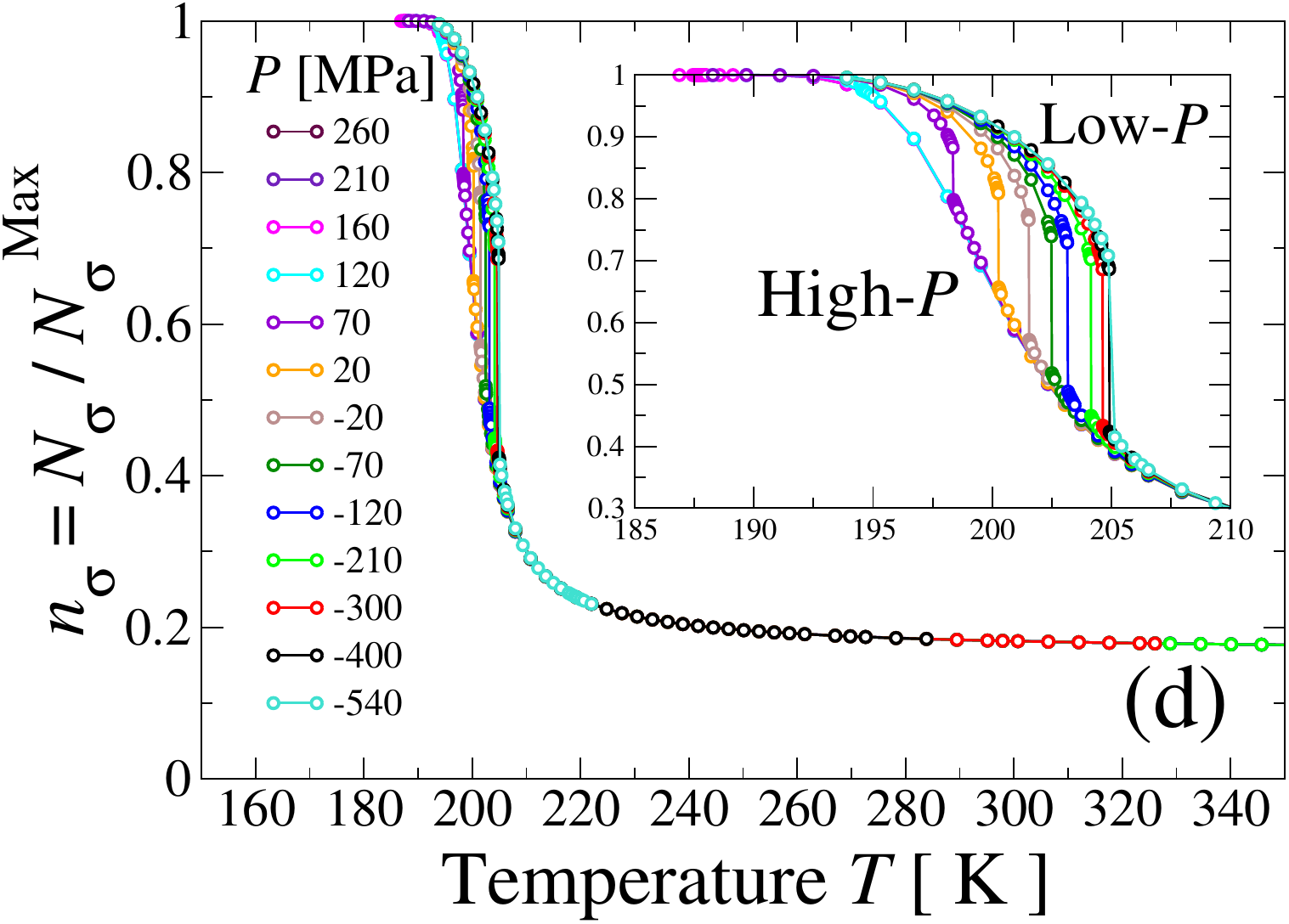}
\caption{{\bf Temperature dependence of FS water thermodynamic quantities along isobars below 260 MPa.} 
In all the panels, circles are MC calculations for $N=$32,768 water molecules, lines are guides to the eyes, 
colors correspond to different pressures indicated in the legend of the bottom panel (from $-540$ MPa to 260 MPa),
red triangles mark temperatures along the LG spinodal,
grey diamonds the TMD.
{\bf (a)} For each $P\leq P_{\rm TMD}^{\rm Max}\simeq 180$ MPa (see also Fig. \ref{fig:density_TMD_limit}), the density $\rho$, expressed in kg/m$^3$, has a maximum  (TMD) and a minimum (TminD, yellow triangles).  
{\bf (b)}  The enthalpy per molecule $h$, expressed in kJ/mol, decreases monotonically for decreasing $T$, with a large variation around a temperature weakly dependent of $P$.
{\bf (c)}  The normalized number of HBs, $n_{\rm HB}$ with  $N_{\rm HB}^{\rm Max}\equiv 2N$, at constant $T$ decreases for increasing $P$. Upon isobaric cooling, it increases for $P$ up to $P_{\rm TMD}^{\rm Max}$ and decreases for higher $P$. The increase is sharper for $P$ approaching $P_{\rm TMD}^{\rm Max}$.
{\bf (d)} The number of cooperative HB bonds $n_\sigma$, normalized with respect to their maximum value $N_{\sigma}^{\rm Max}$, increases at low $T$ with a variation that is smoother at higher $P$. Inset: The detail around the variation reveals that this change correlates with the enthalpy change in (b).
At $T$ larger than the LG spinodal, liquid water transforms into gas, and $\rho$ and $h$ have a discontinuous transition to values below the reported scale.
} 
    \label{fig:supercooled_rho_h_nhb_nsigma}
\end{figure}

At temperatures below the gas-liquid phase transition (LG Spindoal), 
along isobars, FS water displays a temperature of maximum density (TMD) in quantitative agreement with the available experimental data \cite{Coronas-2024}. 
Decreasing the temperature below the TMD leads to a decrease in the isobaric density $\rho$ toward a temperature of minimum density (TminD), with a weak increase below TminD (Fig.~\ref{fig:supercooled_rho_h_nhb_nsigma}a). 
Experiments under confinement, inhibiting water crystallization, show a similar behavior, as recently summarized in Ref.~\cite{Mallamace2024}.

The temperature of the large variation in $\rho$ displays a pronounced $P$-dependence.
Increasing $P$ approaching $P_{\rm TMD}^{\rm Max}\simeq$180 MPa, the jump toward the minimum is sharper compared to low $P$, and the maximum becomes flatter.
Above $P_{\rm TMD}^{\rm Max}$, the density becomes monotonically decreasing for increasing $T$, and the TMD ends \cite{Coronas-2024}, in agreement with the experiments \cite{Mishima:2010fk, Mallamace2024}. The value of $P_{\rm TMD}^{\rm Max}$ can be calculated in the model, in its simplified assumption for the volume Eq.~\ref{eq:Vtot} which implies that the liquid compressibility does not change with $P$ (Fig. \ref{fig:density_TMD_limit}). 

The $P$-dependence of the change in $\rho$ could be apparently consistent with the LLPC scenario at negative $P$ \cite{Tanaka96}.
However, a similar behavior, without any discontinuity, is also predicted by the singularity-free scenario \cite{Stokely2010, Sastry:1996aa, StanleyTeixeira1980}. 

Therefore, we analyze the enthalpy per molecule $h$ and find variations at any $P$ (Fig.~\ref{fig:supercooled_rho_h_nhb_nsigma}b). 
Opposite to density changes, 
the variation in $h$ is sharper at lower $P$ and smoother at higher $P$, with an amplitude of about 5 kJ/mol that is approximately independent of $P$ in the range of pressures we consider.
Also, in contrast with $\rho$, 
the temperature of maximum variation of $h$, between 197~K and 205~K, has a weaker pressure dependence. 
Because $h$ depends both on the density and the HB interactions, the mismatch between the $P$-dependence of enthalpy and density demonstrates that the largest variation in $h$ is controlled by the HB network and its rearrangement. 

On the other hand, also $\rho\equiv V_{\rm Tot}/N$ depends on the number of HBs, $N_{\rm HB}$, Eq.~\ref{eq:Vtot}. For decreasing $T$, $N_{\rm HB}$ rapidly increases at high $P$ up to $P_{\rm TMD}^{\rm Max}$, and displays a smoother increase at low $P$ (Fig.~\ref{fig:supercooled_rho_h_nhb_nsigma}c). 
At constant $T$, $N_{\rm HB}$ decreases for increasing $P$ consistent with the experiments \cite{Mallamace2024}.

At high $T$, $N_{\rm HB}$ converges toward a low average value.  Around atmospheric pressure, this value is close to the lowest possible $\approx$20\%, corresponding to the stochastic probability of having two nearby water molecules with one of their H atoms in between (Appendix~\ref{app:TheFSModel}).

At low temperatures and $P>P_{\rm TMD}^{\rm Max}$, above the TMD, $N_{\rm HB}$ decreases because of the $\rho$-increase that breaks the tetrahedral HB network. This is consistent with the known phenomenon of melting ice by pressurization due to the change of slope of the melting temperature around 200 MPa \cite{Mishima:2010fk}. These high pressures also mark the onset of the ice polymorphism, leading to the formation of interpenetrating HB networks at larger pressures \cite{Lobban:1998aa, Gasser:2021aa}.

At low temperatures below $P_{\rm TMD}^{\rm Max}$,  $N_{\rm HB}$ saturates to two HBs per molecule ($N_{\rm HB}^{\rm Max} \equiv 2N$), i.e., every water molecule is participating in four HBs. However, the changes of $N_{\rm HB}$ at low pressure are not discontinuous as in the enthalpy, showing that the contribution to $h$ coming from HB cooperativity is relevant (Appendix~\ref{app:TheFSModel}).

In particular, $N_\sigma$ is the number of locally cooperative, or tetrahedrally arranged, HBs (Appendix~\ref{app:TheFSModel}). It displays a rapid increase that is sharper at low pressure and smoother for $P > 70$ MPa (Fig.~\ref{fig:supercooled_rho_h_nhb_nsigma}d).
Unlike $N_{\rm HB}$, the change in $N_\sigma$ occurs between 197 K and 205 K, with a weak dependence on $P$, correlating with the variation in $h$.

Therefore, the significant decrease of $h$ is associated with the cooperative contribution, resulting from an extensive structural rearrangement of the HBs towards a more tetrahedral configuration. However, this reorganization implies only a minor change in $\rho$ at low pressure, as it occurs when the number $N_{\rm HB}$ of HBs is almost saturated. 

On the other hand, increasing $P$ toward $P_{\rm TMD}^{\rm Max}$, the formation of a large amount of HBs occurs at low $T$, where the many-body interaction is already relevant. Therefore, at high $P$, the effect of $N_{\rm HB}$ on the density is large and collective, as expected at a critical phase transition.

\subsection{Isobaric specific heat.} 

\begin{figure}
    \centering
    \includegraphics[scale=0.3]{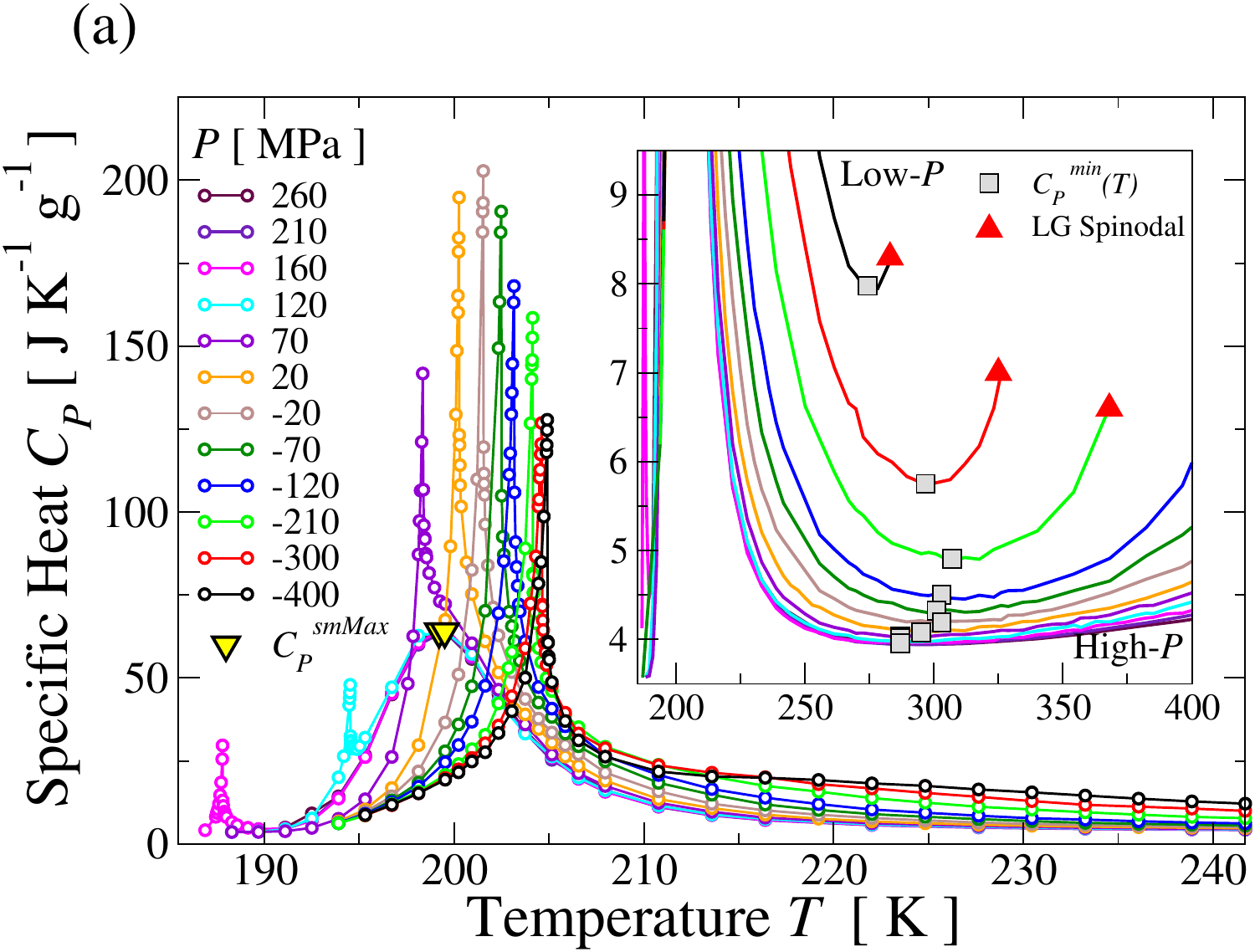}
    \includegraphics[scale=0.3]{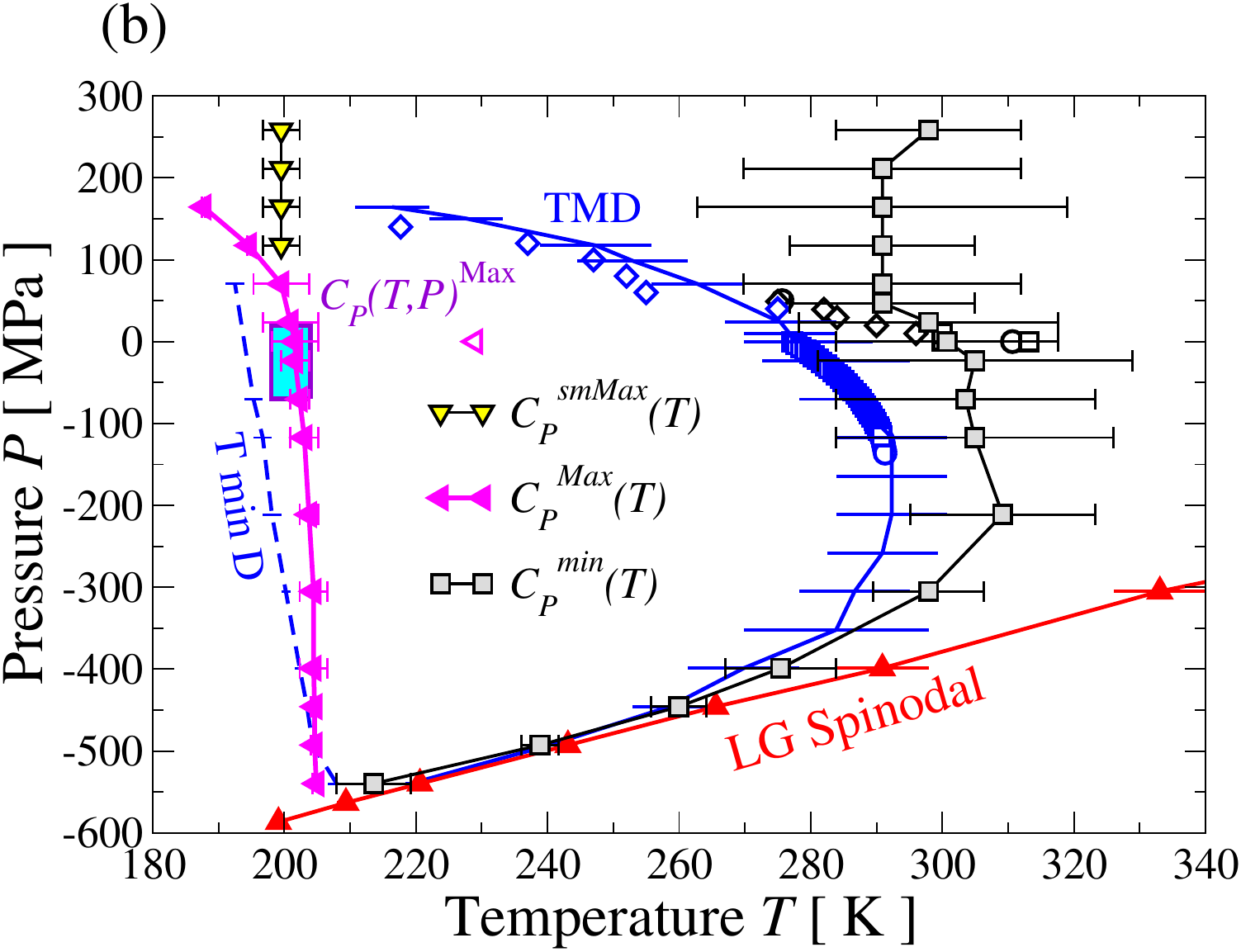}
\caption{{\bf The isobaric specific heat has maxima at low temperatures with apparent divergence below 200 MPa.}
In all panels, the MC calculations are for $N=$32,768 water molecules,  red triangles are points on the LG spinodal,
grey squares mark specific heat minima $C_P^{\rm min}(P)$, 
and
yellow triangles mark smooth maxima $C_P^{\rm smMax}(P)$.
{\bf (a)}  $C_P$, in J/(gK), as a function of $T$, in K, along isobars from $-540$ MPa to 260 MPa (with colors as indicated in the legend).
For clarity, we show only a selection of results from the fluctuation-dissipation theorem.
Circles are the calculations, and lines are guides to the eyes. 
Sharp maxima, $C_P^{\rm Max}(P)$, occur at any $P$, achieving higher values between $-70$ MPa and 20 MPa.
For $P\geq 120$ MPa, we find also smooth maxima, $C_P^{\rm smMax}(P)$, almost $P$-independent.
Inset: The details of the same data, represented as lines for clarity, at high $T$ around the $C_P^{\rm min}(P)$.
{\bf (b)} In the $(T, P)$ thermodynamic plane,
the locus of $C_P^{\rm min}(P)$ converges asymptotically to the LG spinodal, at $P<0$, before turning into the locus of $C_P^{\rm Max}(P)$ (magenta left triangles), where the TMD (blue solid) line turns into the TminD (blue dashed) line.
At high $P$, the locus of $C_P^{\rm smMax}(P)$ departs from that of $C_P^{\rm Max}(P)$ near the state points where $C_P$ apparently diverges (cyan region). 
The TMD line and the $C_P^{\rm min}(P)$ locus are in quantitative agreement within the statistical error with the experimental data above the homogenous nucleation temperature $T_h$~\cite{Mallamace2024} (not shown) for 
the TMD (blue diamonds \cite{Mishima:2010fk}; blue squares \cite{Pallares2014}; blue circles  \cite{Holten:2017um}),  
and
$C_P^{\rm min}(P)$ (black diamonds \cite{Stepanov2014}; black circles \cite{LinTrusler2012}; black squares \cite{CRCHandbook2020}), respectively.
Below $T_h$, the agreement with the estimated
$C_P^{\rm Max}(P)$ at atmospheric pressure (magenta open triangle \cite{doi:10.1073/pnas.2018379118}) is only qualitative, calling for further investigation.
}
    \label{fig:Supercooled_CP}
\end{figure}

To test whether the observed thermodynamic behavior is consistent with criticality, we calculate the response functions $C_P$, $K_T$, and $\alpha_P$ and study if they diverge at the hypothesized LLCP.
We find sharp maxima in $C_P \equiv N \left(\langle h^2\rangle -
\langle h \rangle^2\right)/k_BT^2$ at any pressure $P$ and low $T$ 
with an apparent divergence at $-70$~MPa$\leq P\leq 20$~MPa within our numerical resolution (Fig.~\ref{fig:Supercooled_CP}a).
 
For pressures below 20 MPa, the locus of the maxima of $C_P$ depends weakly on $P$, occurring around 200 K, in the range of $T$ where $h$ and $N_\sigma$ have large changes
(Fig.~\ref{fig:Supercooled_CP}b).  
For pressures $P>20$ MPa, the sharp maxima decrease in intensity and move toward lower temperatures. This behavior would be consistent with 
an LLPT line with a negative slope in $(T, P)$ ending in an LLCP where $C_P(T)$ apparently diverges (Fig.~\ref{fig:Supercooled_CP}b).

Between the TMD line and the LG spinodal, $C_P$ has isobaric minima (Fig.~\ref{fig:Supercooled_CP}a, inset). 
Between 0 and 50 MPa, the minima agree, within the statistical error, with the experimental data \cite{Coronas-2024}.

As $P$ decreases, the locus $C_P^{\rm min}(P)$ approaches the TMD line and the LG spinodal, asymptotically (Fig.~\ref{fig:Supercooled_CP}b).
The TMD line and the LG spinodal do not intersect, which is consistent with the monotonicity of the spinodal pressure versus temperature. The LG spinodal should otherwise be reentrant in case of intersection with the TMD \cite{Speedy:1982aa}.

\begin{figure}[!ht]
    \centering
    \includegraphics[scale=0.3]{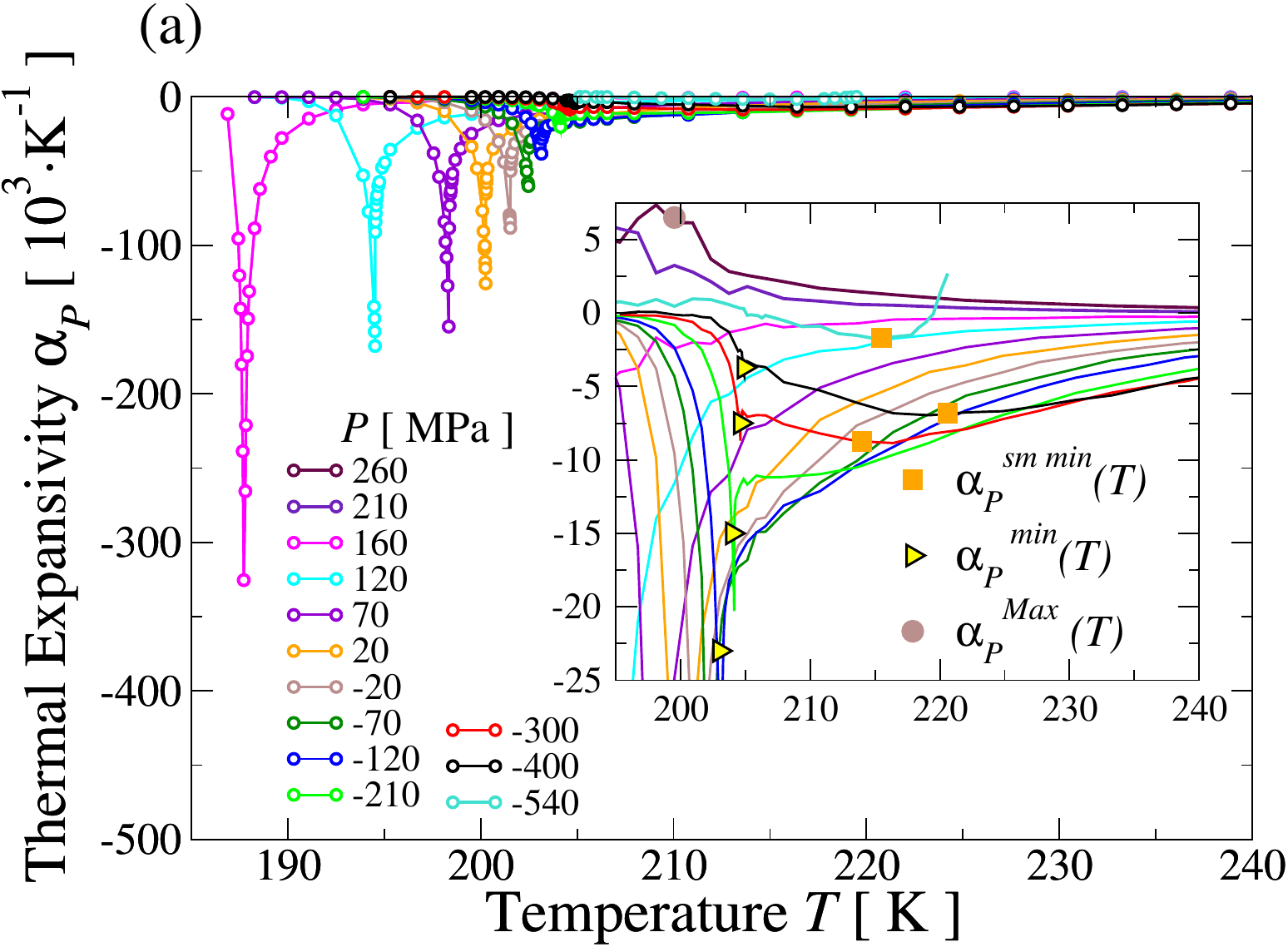}
    \includegraphics[scale=0.3]{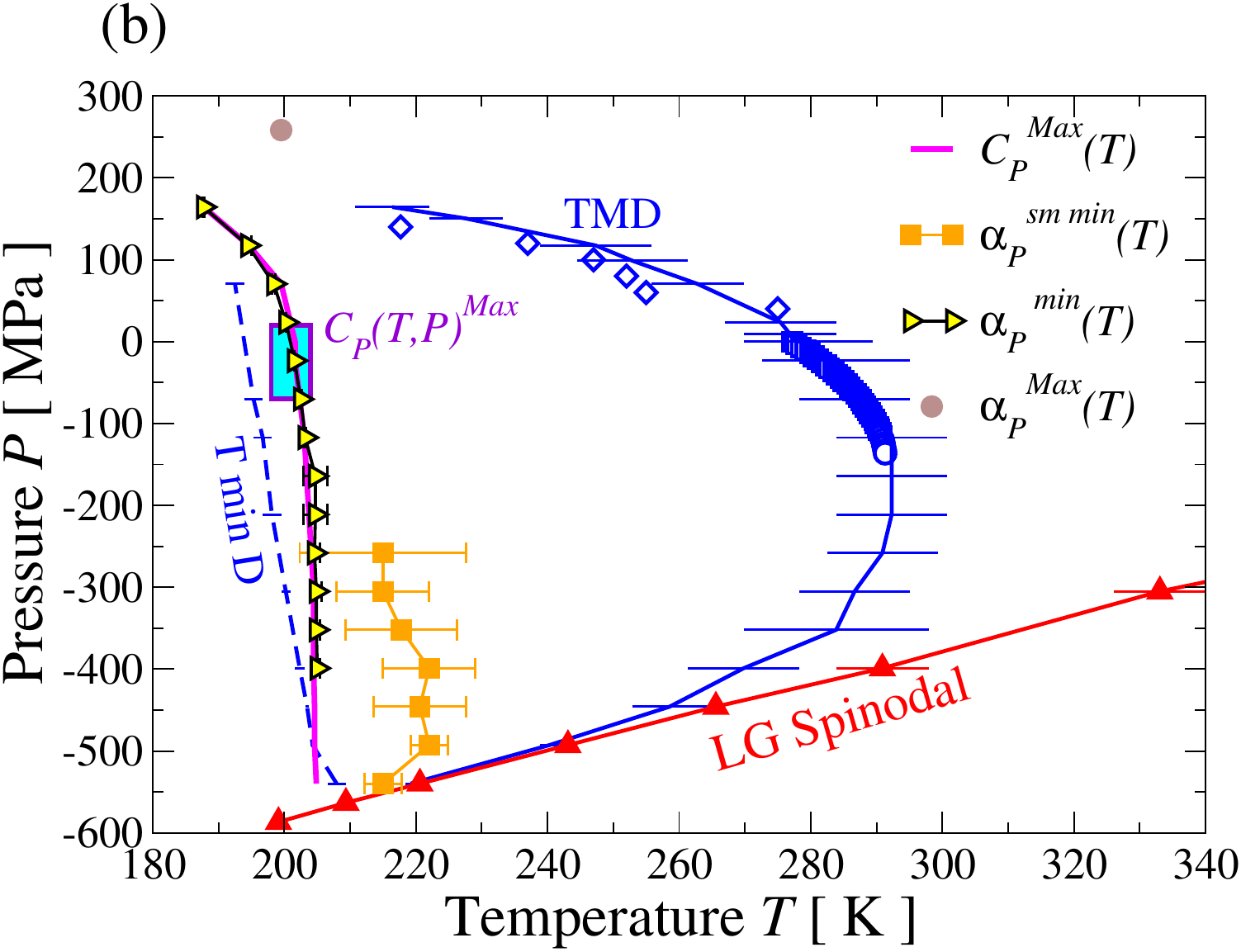}
\caption{{\bf The isobaric thermal expansivity grows negative below the TMD line with extrema along the 
$C_P^{\rm Max}(P)$ locus.}
In all panels, the MC calculations are for $N=$32,768,
yellow triangles mark sharp minima $\alpha_P^{\rm min}(P)$, 
orange squares mark smooth minima $\alpha_P^{\rm sm~min}(P)$,
brown circle marks a maxima $\alpha_P^{\rm Max}(P)$ at 260 MPa.
{\bf (a)} $\alpha_P$, in K$^{-1}$, as a function of $T$, in K, along isobars from $-540$ MPa to 260 MPa (with colors as indicated in the legend).  For clarity, we show only the results from the fluctuation-dissipation theorem (circles) for a selection of the calculated isobars.
Lines are guides to the eyes. 
The expansivity has minima, $\alpha_P^{\rm min}(P)$, that are sharp and strong at high $P>0$. They become weaker as $P$ decreases. 
Inset: 
The same data as in the main panel, represented as lines for clarity, around the $\alpha_P^{\rm sm~min}(P)$, for $T$ above the $\alpha_P^{\rm min}(P)$. The sharp minima disappear for $P < -400$ MPa, while, for $P \leq -250$ MPa, the expansivity develops the smooth minima at higher-$T$.
We find a maxima $\alpha_P^{\rm Max}(P)$ at 260 MPa. 
{\bf (b)} 
The locus of $\alpha_P^{\rm min}(P)$ follows the locus of $C_P^{\rm Max}(P)$ (magenta line), crossing the state points where $C_P$ apparently diverges (cyan region). The locus of $\alpha_P^{\rm sm~min}(P)$ converges at low $P$ toward the $C_P^{\rm Max}(P)$ line and crosses the TMD (blue solid) line where it turns into the TminD (blue dashed) line, tangent to the LG spinodal (red line). Experimental data for the TMD line are as in Fig.\ref{fig:Supercooled_CP}b.}
    \label{fig:Supercooled_AlphaP}
\end{figure}

\begin{figure}[!ht]
    \centering
    \includegraphics[scale=0.3]{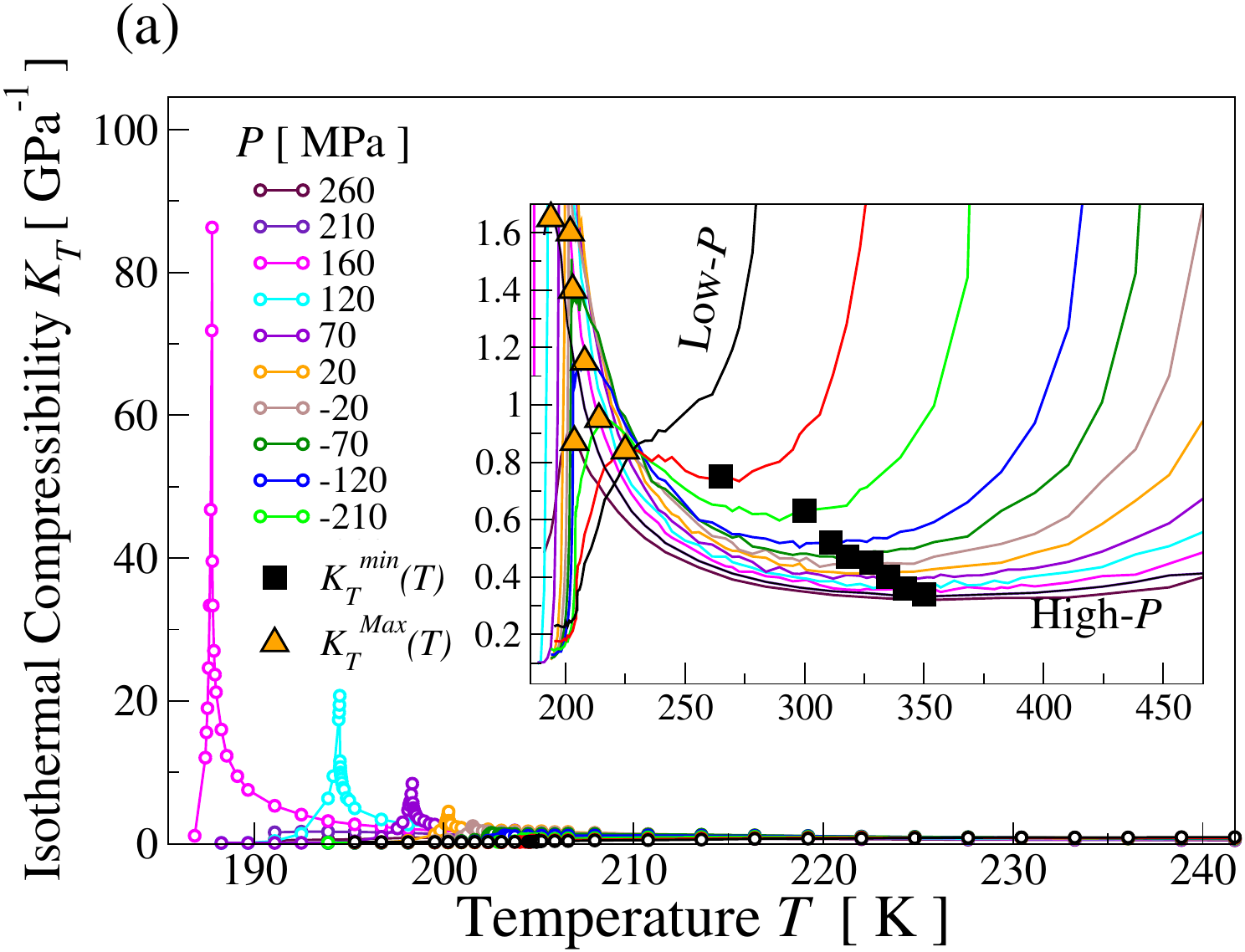}
    \includegraphics[scale=0.3]{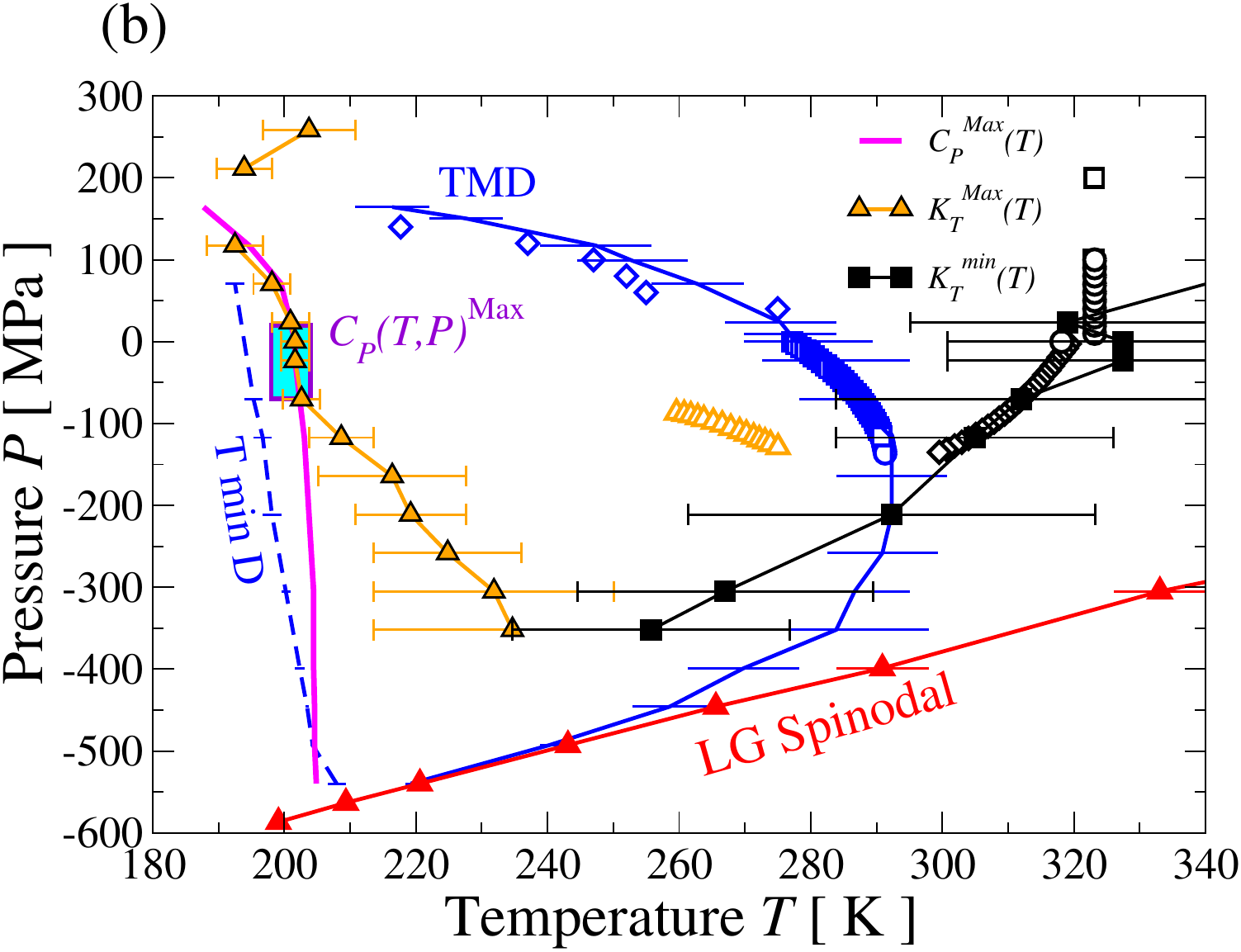}
\caption{{\bf The isothermal compressibility has maxima below the TMD line merging the $C_P^{\rm Max}(P)$ line where $C_P$ apparently diverges.}
In all panels, the MC calculations are for $N=$32,768, 
brown upper triangles and black squares mark maxima $K_T^{\rm Max}(P)$ and minima $K_T^{\rm min}(P)$, respectively.
{\bf (a)}  $K_T$, in GPa$^{-1}$, as a function of $T$, in K, along isobars
from $-540$ MPa to 260 MPa (with colors as indicated in the legend).
For clarity, we show only the results (circles) from the fluctuation-dissipation theorem for a selection of the calculated isobars.
Lines are guides to the eyes. 
The maxima $K_T^{\rm Max}(P)$ rapidly decrease upon decreasing $P<200$ MPa. Inset:
The same data as in the main panel, represented as lines for clarity, showing the convergence of the $K_T^{\rm Max}(P)$ and the minima $K_T^{\rm min}(P)$. We show also the two maxima above 200 MPa at $\simeq 187$ K and 202 K.
{\bf (b)}  
The $K_T^{\rm Max}(P)$ line converges to the (cyan) region where $C_P$ apparently diverges and merges with the $C_P^{\rm Max}(P)$ (magenta) line.
The locus of $K_T^{\rm min}(P)$ crosses the TMD (blue) line in its turning point (point of maximum slope), in quantitative agreement with the available experimental data above the TMD (black empty squares \cite{TerMinassian1981}; black circles \cite{FineMillero1973}; black diamonds \cite{Holten:2017um}). The agreement with the experimental data for $K_T^{\rm Max}(P)$ in stretched water below the TMD (orange empty triangles \cite{Holten:2017um}) is only qualitative and should be further investigated.
We find $K_T^{\rm Max}(P)$ also at pressures above 200 MPa.}
    \label{fig:Supercooled_KT}
\end{figure}

The locus $C_P^{\rm min}(P)$ crosses the TMD line where the minima in $C_P$ turn into maxima $C_P^{\rm Max}(P)$ and, at the same time, the TMD line turns into the TminD line, as expected for thermodynamic consistency. This is due to the relation \cite{Bianco2014}
\begin{equation}\label{rel2}
 \left(\frac{\partial C_P}{\partial P}\right)_T=T\left(\frac{\partial P}{\partial T}\right)_{\rm TMD}\left(\frac{\partial^2V}{\partial P\partial T}\right)_{\rm TMD}
\end{equation} 
indicating that the TMD line has a turning point in the $(T, P)$ plane, 
$(\partial P/\partial T)_{\rm TMD}=0$, when it intersects 
 the locus of $C_P$ extrema, $(\partial C_P/\partial P)_T=0$  
\cite{Poole2005, Bianco2014, Holten:2017um}.

At $P>70$ MPa, we find that $C_P$ develops a smooth isobaric maximum $C_P^{\rm smMax}(P)$ at intermediate $T$, between the loci of $C_P^{\rm Max}(P)$ and  $C_P^{\rm min}(P)$ (Fig.~\ref{fig:Supercooled_CP}). 
The locus $C_P^{\rm smMax}(P)$ is necessary to maintain thermodynamic consistency as occurs for other thermodynamic response functions \cite{Buldyrev2009}. As $\rho$ increases, water loses its anomalies \cite{Errington2001}. As density always increases with increasing pressure, a similar relationship holds with $P$. Hence, the locus $C_P^{\rm min}(P)$ must have a turning point at high $P$ merging
 into a locus of maxima of $C_P$ occurring at lower $T$.

However, this locus of maxima of $C_P$ cannot be the same as $C_P^{\rm Max}(P)$, as the latter should follow the LLPT that is expected to flatten out at high $P$ \cite{Mishima1998, Mallamace2024}. Hence, although we do not reach the high-pressure region where it occurs, $C_P^{\rm smMax}(P)$ must 
merge with the $C_P^{\rm min}(P)$ line, enclosing the anomalous region of $C_P$, to preserve the thermodynamic consistency. The study of this merging point goes beyond the scope of the current work.

The two $C_P$ maxima observed here are qualitatively different from those discovered by Mazza et al. for the FS monolayer \cite{Mazza2012}. Consistent with experiments for a water monolayer \cite{Mazza2011}, Mazza et al. found the two $C_P$ maxima also at low pressures, approaching the LG spinodal. Here, instead, we find two maxima only at pressures above the possible critical region, as in experiments of confined water \cite{Cupane2014, Mallamace2020} and atomistic simulations of bulk water \cite{Eltareb:2022wk, Gonzalez2016, Holten2014}.

\subsection{Isobaric thermal expansivity.} 

Next, we calculate the thermal expansivity,  $\alpha_P \equiv 
(\langle VhN\rangle - \langle V \rangle\langle hN\rangle)/(k_B  
T^2\langle V \rangle)$, along isobars (Fig.~\ref{fig:Supercooled_AlphaP}a). We observe sharp minima, $\alpha_P^{\rm min}(P)$, 
along the $C_P^{\rm Max}(P)$ line,
with decreasing intensity as the $P$ reduces  (Fig.~\ref{fig:Supercooled_AlphaP}b). 
At $-210$ MPa, $\alpha_P$ develops a shoulder at temperatures above the minimum (Fig.~\ref{fig:Supercooled_AlphaP}a, inset). The shoulder becomes a smooth minimum $\alpha_P^{\rm sm~min}(P)$ as the pressure decreases. 
For $-400 {\rm~MPa}\leq P\leq -250 {\rm~MPa}$, $\alpha_P$ has a primary minimum, $\alpha_P^{\rm sm~min}(P)$, at higher $T$ and a secondary 
minimum, $\alpha_P^{\rm min}(P)$, at lower $T$.

The fact that both $\alpha_P$, proportional to the cross-fluctuations of entropy and volume \cite{FS2007},  and $C_P$, proportional to the fluctuations of entropy \cite{FS2007}, have extrema along the same locus in the plane $(T, P)$ indicates that at these specific state points there is the largest rearrangement of the HB network towards a more tetrahedral ordering.  This is in agreement with the $N_\sigma$ large variation observed in Fig.\ref{fig:supercooled_rho_h_nhb_nsigma}(d).
Hence, the sharp minima in $\alpha_P$ corresponds to a maximum in the structural rearrangement of the HB network.

The smooth minima in $\alpha_P$ at negative $P$ indicate that the HB ordering can occur progressively at intermediate $T$. The $\alpha_P^{\rm sm~min}(P)$ line correlates in $T$ with the largest variation of the gradual increase of $N_{\rm HB}$ at low $P$ (Fig.\ref{fig:supercooled_rho_h_nhb_nsigma}c). 

At higher and positive $P$, the energy gain of the HBs can overcome their high enthalpy cost, due to the local volume increase, only at low enough $T$, inducing a merging, within the error bar, of the smooth and the sharp minima. This should occur at the $C_P^{\rm Max}(P)$ line, where $N_\sigma$ and $N_{\rm HB}$ increase together at high $P$, although it is only suggested by our calculations. 

Also, all these observations are consistent with the mean-field results showing that $\alpha_P$ is proportional to the isobaric $T$-derivative of the probability of forming HBs, apart from a term that is relevant only near the LG spinodal \cite{FS2007}. Hence, it correlates with the HB-network structural changes marked by the derivatives of $N_\sigma$ and $N_{\rm HB}$. 

It is interesting to note that our results indicate that the positions of the two minima, $\alpha_P^{\rm sm~min}(P)$ and $\alpha_P^{\rm min}(P)$, become identical when they approach the LG spinodal. The position of $\alpha_P^{\rm sm~min}(P)$ becomes tangential to both the LG spinodal and the TMD line before turning towards the $\alpha_P^{\rm min}(P)$ position where the TMD line and the TminD line merge  (Fig.~\ref{fig:Supercooled_AlphaP}b). This is consistent with the thermodynamic relation 
\begin{equation}\label{rel4}
\left(\frac{\partial \alpha_P}{\partial T}\right)_{P,~\rm TMD}=-\frac{1}{V}\left(\frac{\partial P}{\partial T}\right)_{\rm TMD}\left(\frac{\partial^2V}{\partial P\partial T} \right)_{\rm TMD},
\end{equation} 
showing that $\alpha_P$ has zero $T$-derivative, as in a minima, if it crosses the TMD line where the slope is zero, i.e., where it turns into the TminD line.

In the same way, a similar relationship would also hold for a maximum in $\alpha_P$. We can observe a potential maximum at -540 MPa around 205 K (Inset Fig.\ref{fig:Supercooled_AlphaP}a), and a definite maximum $\alpha_P^{\rm Max}(P)$ at approximately the same temperature but at a very high pressure (260 MPa). This suggests that, as observed for the other response functions, the locations of the extremes should trend towards high pressure and high temperature and then converge towards the region where $C_P$ seems to diverge, forming a closed region in the $(T, P)$ plane. Further investigation beyond the scope of the present work will be required to understand this feature.

\subsection{Isothermal compressibility.} 

Finally, we calculate the isothermal compressibility, 
$K_T \equiv \langle V^2\rangle/k_BT\langle V\rangle$, along isobars (Fig.~\ref{fig:Supercooled_KT}a). We observe that $K_T$ has maxima, $K_T^{\rm Max}(P)$, that in the $(T,P)$ plane converge toward the extrema of $C_P$ and $\alpha_P$ (Fig.~\ref{fig:Supercooled_KT}b). All the extrema of the three response functions merge in the region where $C_P$ apparently diverges and follow each other at higher $P$, as it would be expected along a first-order LLPT with a negative slope in the $(T, P)$ plane ending in an LLCP. 

As the pressure decreases, the maxima $K_T^{\rm Max}(P)$ decrease and turn into minima, $K_T^{\rm min}(P)$. These minima occur at increasing temperatures with increasing pressure above $-350$ MPa and cross the TMD line at its turning point. As discussed in Refs. \cite{Poole2005, Bianco2014, Holten:2017um}, this is a consequence of the thermodynamic relation \cite{Bianco2014}
\begin{equation}\label{rel1}
\left(\dfrac{\partial K_T}{\partial T}\right)_{P,~\rm TMD}=
\dfrac{1}{V} \dfrac{(\partial^2 V/\partial T^2)_{\rm TMD}}{ (\partial
  P/\partial T)_{\rm TMD}} 
\end{equation} 
indicating that, at the TMD line,  $K_T$ has a zero $T$-derivative along isobars when the TMD line has an infinite slope.
At pressures above 200 MPa, we find another branch of the locus of $K_T^{\rm Max}(P)$, as expected for thermodynamic consistency~\cite{Luo2014, Buldyrev:2015uq}.

We observe that our calculations recover, within the error bars, the thermodynamic relation 
\begin{equation}\label{alpha_cross}
\left(\dfrac{\partial \alpha_P}{\partial P}\right)_{T}=
- \left(\dfrac{\partial K_T}{\partial T}\right)_P
\end{equation} 
indicating that the extrema of $K_T$ along isobars correspond to state points where  $\alpha_P$ is constant along isotherms. Consequently, we find that $\alpha_P$, calculated at close pressures, has crossing temperatures coinciding with the loci of extrema of $K_T$ (Fig. \ref{fig:alpha_crossing}).

\begin{figure}[!ht]
    \centering
    \includegraphics[scale=0.3]{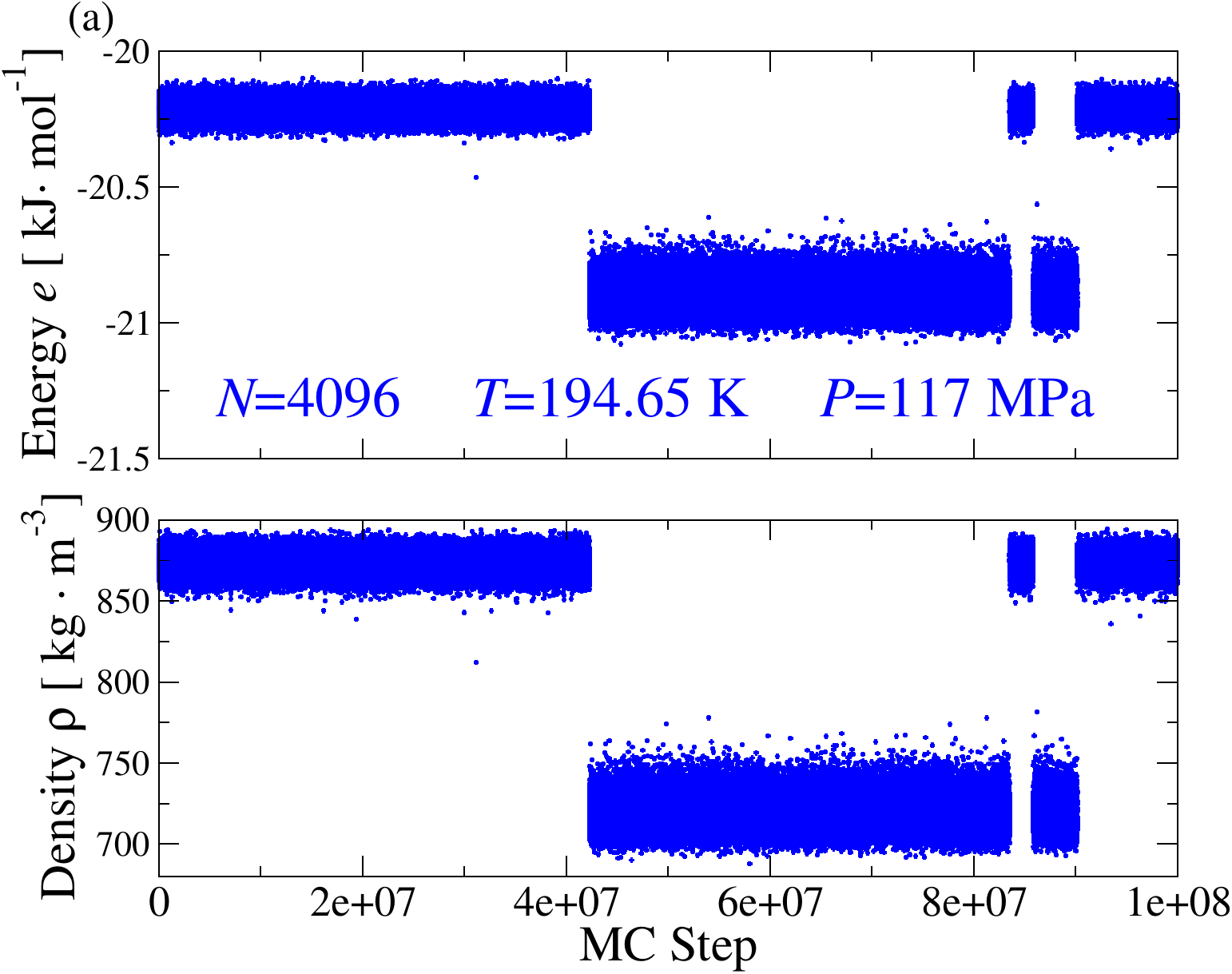}
    \includegraphics[scale=0.29]{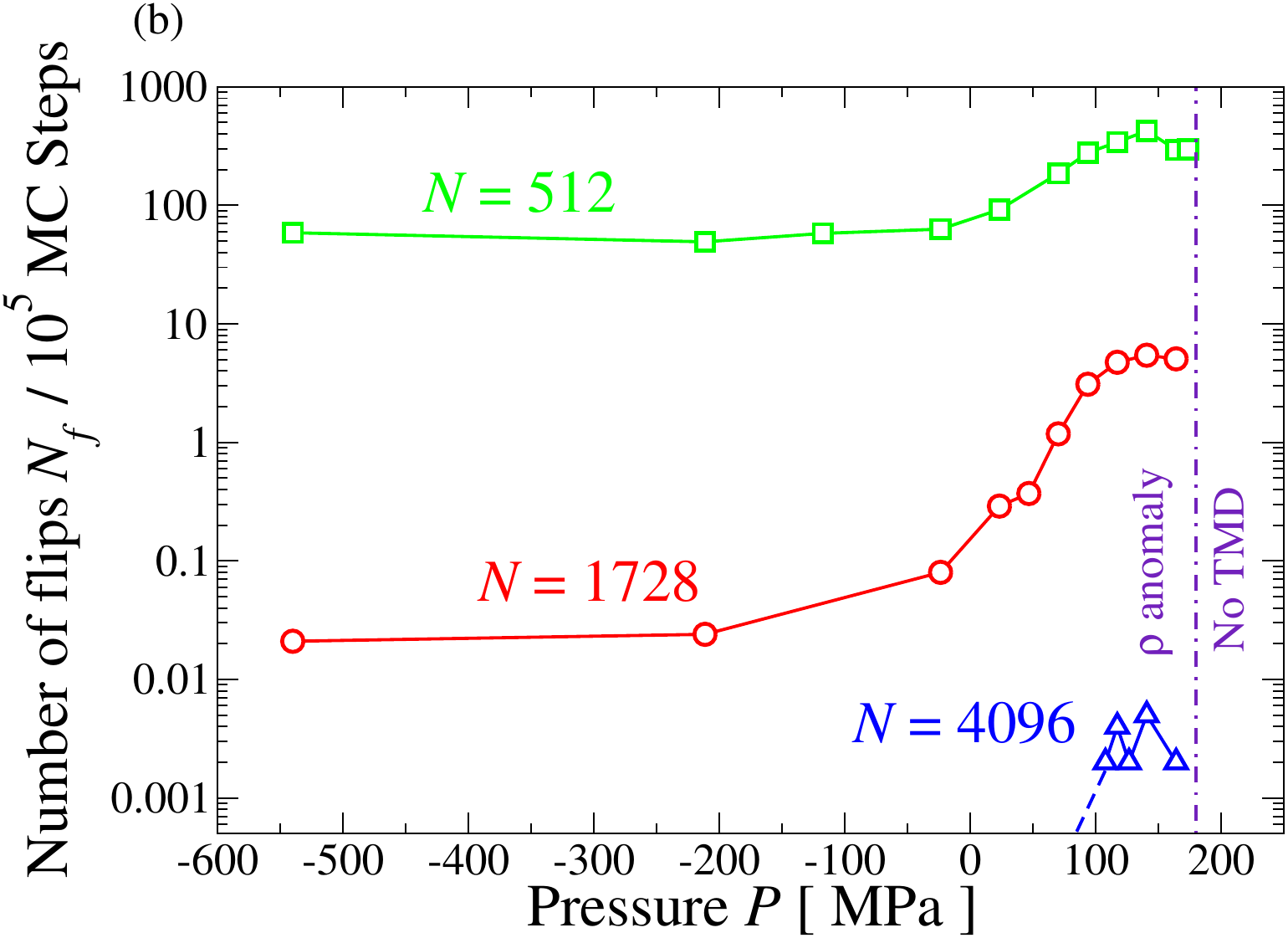}
\caption{{\bf Correlated flipping in energy and density between HDL-like and LDL-like states.}
{\bf (a)} Sequence of (top panel) molar energy $e$, in kJ/mol, and (bottom panel) density $\rho$, in kg/m$^3$ for the configurations generated by the MC algorithm for a system with $N=$4,096 water molecules at $T=194.65$ K, and $P=117$ MPa.
We observe four correlated flips between states with high $(e, \rho)$ and states with low $(e, \rho)$, as between HDL and LDL, respectively. 
{\bf (b)} Number $N_f$ of flips, within $10^5$ MC steps, between  high-$(e, \rho)$ and  low-$(e, \rho)$ states at fixed $N$, for $N=$512 (green squares), 1,728 (red circles), and 4,096 (blue triangles), as a function of $P$, in MPa, along the $(T, P)$ locus where the maxima in $C_P$, $\alpha_P$, and $K_T$ converge. Continuous lines are guides for the eyes.
For $N=$4,096 water molecules, we do not observe any switching below 110 MPa within $10^8$ MC steps (blue dashed line).
The purple dot-dashed line indicates the pressure $P_{\rm TMD}^{\rm Max}\simeq 180$~MPa  above which the model displays no density anomaly (maximum $P$ of the TMD line, Fig. \ref{fig:density_TMD_limit}), consistent with the experiments \cite{Mishima:2010fk, Mallamace2024}.}
\label{fig:Supercooled_timeseires_LLCP}
\end{figure}

\subsection{\label{sec:ResultsFiniteSizeAnalysis}Finite size analysis for the LLPT and the LLCP}

The results in the previous sections are consistent with an LDL-HDL phase separation in finite systems.  We perform a finite-size scale analysis to demonstrate the occurrence of the LLCP in the thermodynamic limit and estimate its universality class.

The HDL-like and LDL-like states have high molar energy $e$ and $\rho$, and low $e$ and $\rho$, respectively.
Consistently, we observe correlated switching between states with low and high values of $e$ and $\rho$ in our MC calculations (Fig.\ref{fig:Supercooled_timeseires_LLCP}a). 
These flips are consistent with a bimodal joint probability density distribution $Q(e,\rho)$, as in a phase coexistence.

At any number $N$ of water molecules, the number $N_f$ of flips reaches a maximum around 150 MPa along the locus of maxima in $C_P$, $\alpha_P$, and $K_T$. This pressure is close to the upper limit $P_{\rm TMD}^{\rm Max}$ beyond which water has no TMD both in our calculations (Fig. \ref{fig:density_TMD_limit}) and experiments \cite{Mishima:2010fk, Mallamace2024} (Fig.\ref{fig:Supercooled_timeseires_LLCP}b). 

As we increase $N$ from 512 to 4,096, the value of $N_f(P)$ decreases by five orders of magnitude and eventually vanishes below 110 MPa within $10^8$  MC steps (see Table \ref{table:simulation_times} ). However, the flips still occur even for the largest $N$ when $P\geq 110$ MPa. This is consistent with the expected behavior for state points near phase transitions in systems with a size smaller than the correlation length of the o. p. fluctuations. However, finite-size scaling theory allows one to extrapolate $N$-dependent critical parameters to the $N\to\infty$ limit near a critical point.

\begin{figure}
    \centering
    \includegraphics[scale=0.4]{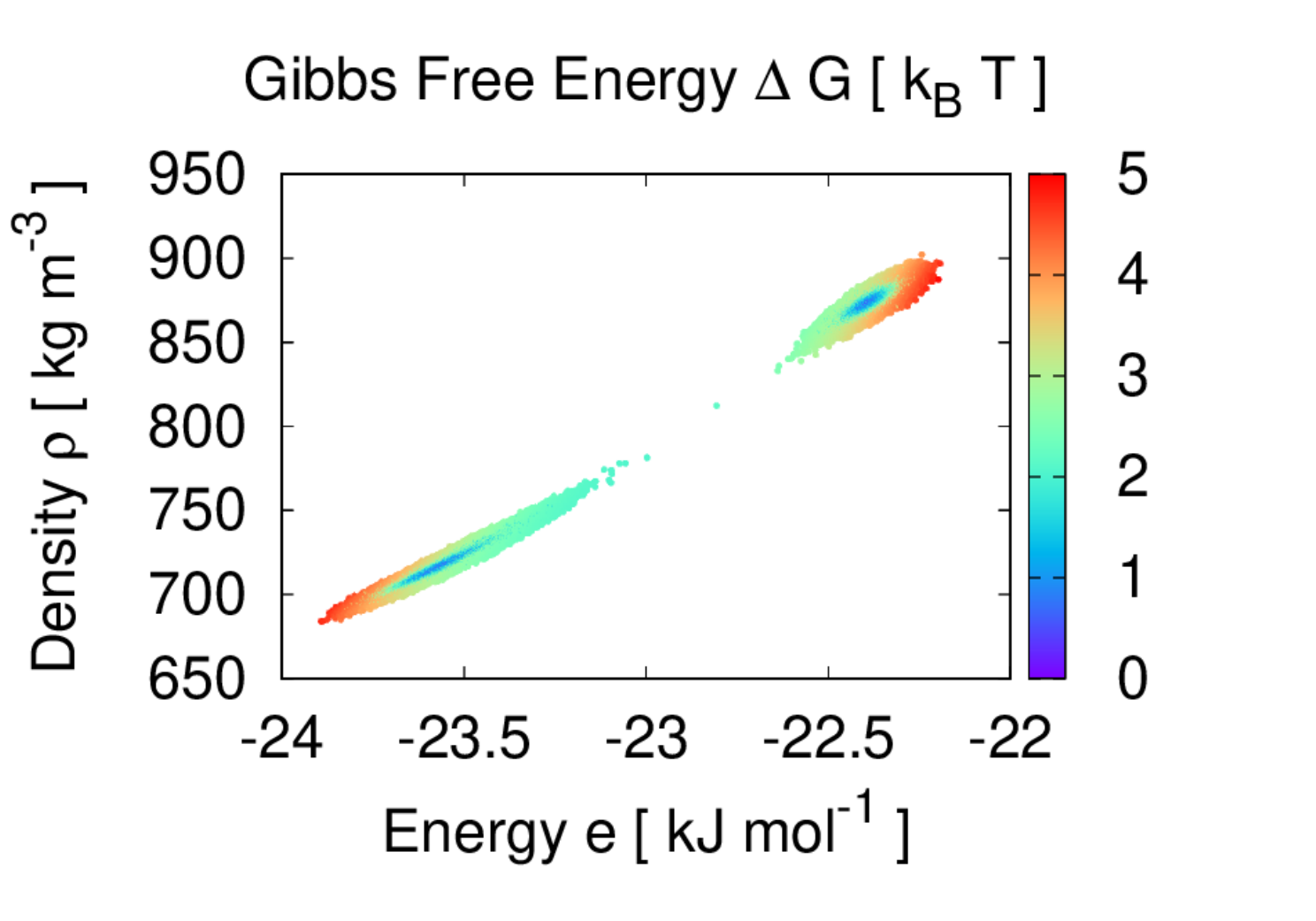}
\caption{{\bf Gibbs free energy difference near the LLCP estimated for a finite system.} $\Delta G$, in units of $k_BT$ (colored-bar scale), for size $N=$4,096, as calculated with the histogram reweighting method, in the $(e, \rho)$ plane, where the molar energy $e$ is expressed in kJ/mol units and the density $\rho$ in kg/m$^3$. The system samples with larger probability the two (blue) regions corresponding to LDL-like and HDL-like states at low-$(e, \rho)$ and high-$(e, \rho)$, respectively.}
    \label{fig:HR}
\end{figure}

To calculate the $N$-dependent critical parameters, we resort to
the histogram reweighting method (section \ref{sec:estimationLLCP}). 
We estimate the free energy landscape $\Delta G(e,\rho) /k_BT = -\log(Q(e,\rho))$ near the LLCP for the largest system size (Fig.~\ref{fig:HR}).
Our analysis reveals two basins of attraction, one associated with a low-$(e, \rho)$ state and the other with a high-$(e, \rho)$ state, as expected at the LDL-HDL coexistence. These two basins are separated by a free-energy barrier of approximately $2 k_BT$, which thermal fluctuations can easily overcome, consistent with the behavior near a critical point.

\begin{figure}
    \centering
    \includegraphics[scale=0.30]{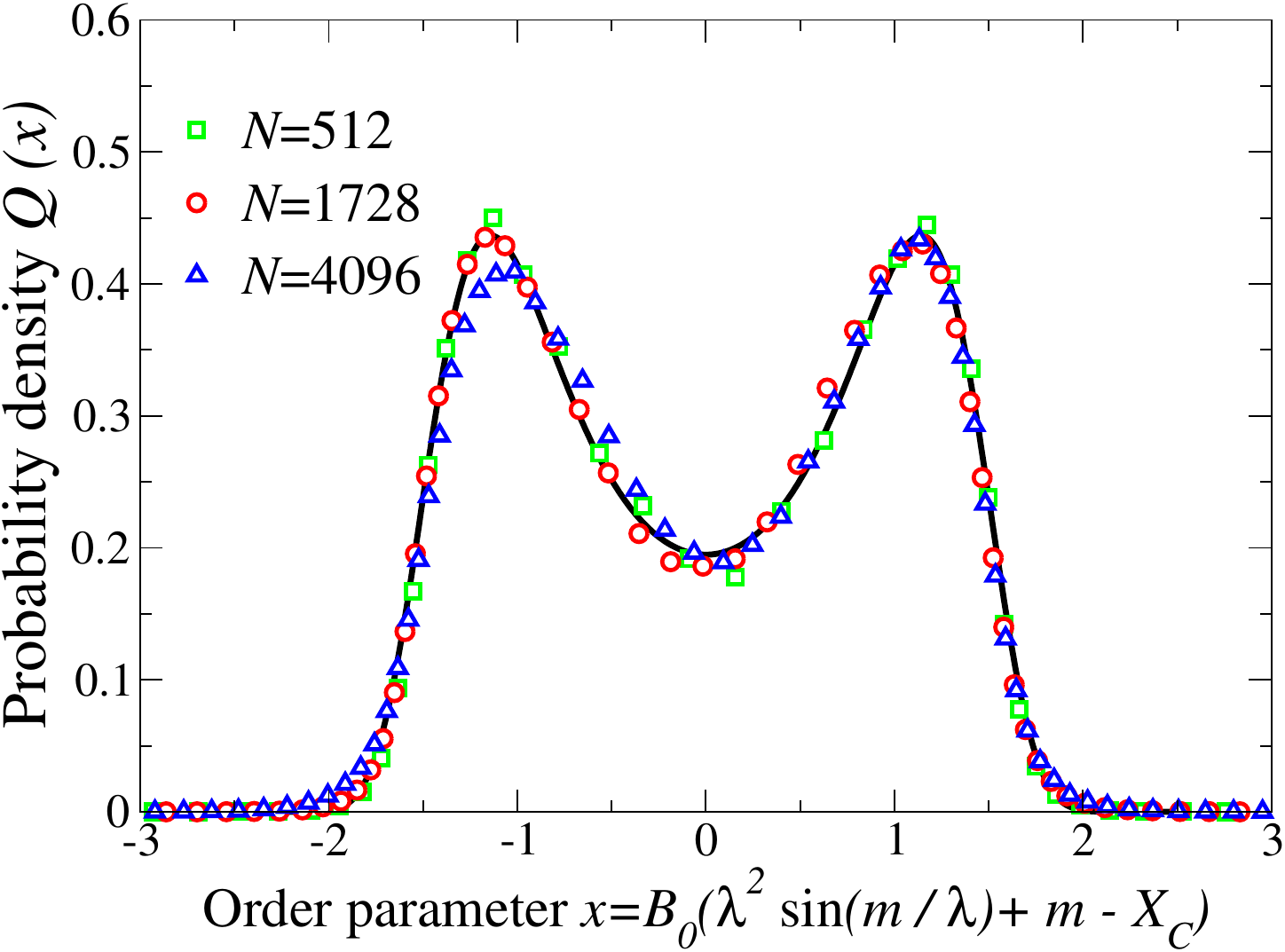}
\caption{{\bf The water LLCP belongs to the 3D Ising universality class.}
The rescaled o. p. $x(s)$, as defined in Appendix~\ref{app:OrderParameter}, at the LLCP follows the 3D Ising critical probability density distribution $Q_3$ (continuous black line) in the thermodynamic limit. Symbols, as indicated in the legend, are for systems with $N=$512 (green squares), 1,728 (red circles), and 4,096 (blue triangles) water molecules. The critical parameters are reported in Table~\ref{table:critical_parameters}.}
    \label{fig:Ising3D}
\end{figure}

To localize the LLCP accurately and analyze its universality class in the thermodynamics limit, we estimate the mixing parameter $s$, which defines the o. p. $M(s)$ and its rescaled version $x(s)$, and the size-dependent critical temperature and pressure $\{T_C(N), P_C(N)\}$,  as described in section \ref{sec:estimationLLCP}. Our results show that the
$x(s)$ at the size-dependent LLCP follows the probability density distribution $Q_3$ of the 3D Ising critical point (Fig.~\ref{fig:Ising3D}). Therefore, we conclude that the LLCP belongs to the 3D Ising universality class in the thermodynamic limit.

In Table~\ref{table:critical_parameters}, we indicate the parameters adopted in Fig.~\ref{fig:Ising3D} for each $N$, finding strong finite-size effects. For $N=$512, we find fluctuations of $x$ that follow $Q_3$ for a wide range of $P$ and a limited range of $T$ along the $C_P^{\rm Max}(P)$ line, as discussed in the next section. As $N$ increases, the range of $P$ at which $Q(x)$ is compatible with $Q_3$ becomes narrower, indicating that, for larger $N$, the region with critical fluctuations is smaller.

\begin{figure}[!ht]
\includegraphics[scale=0.34]{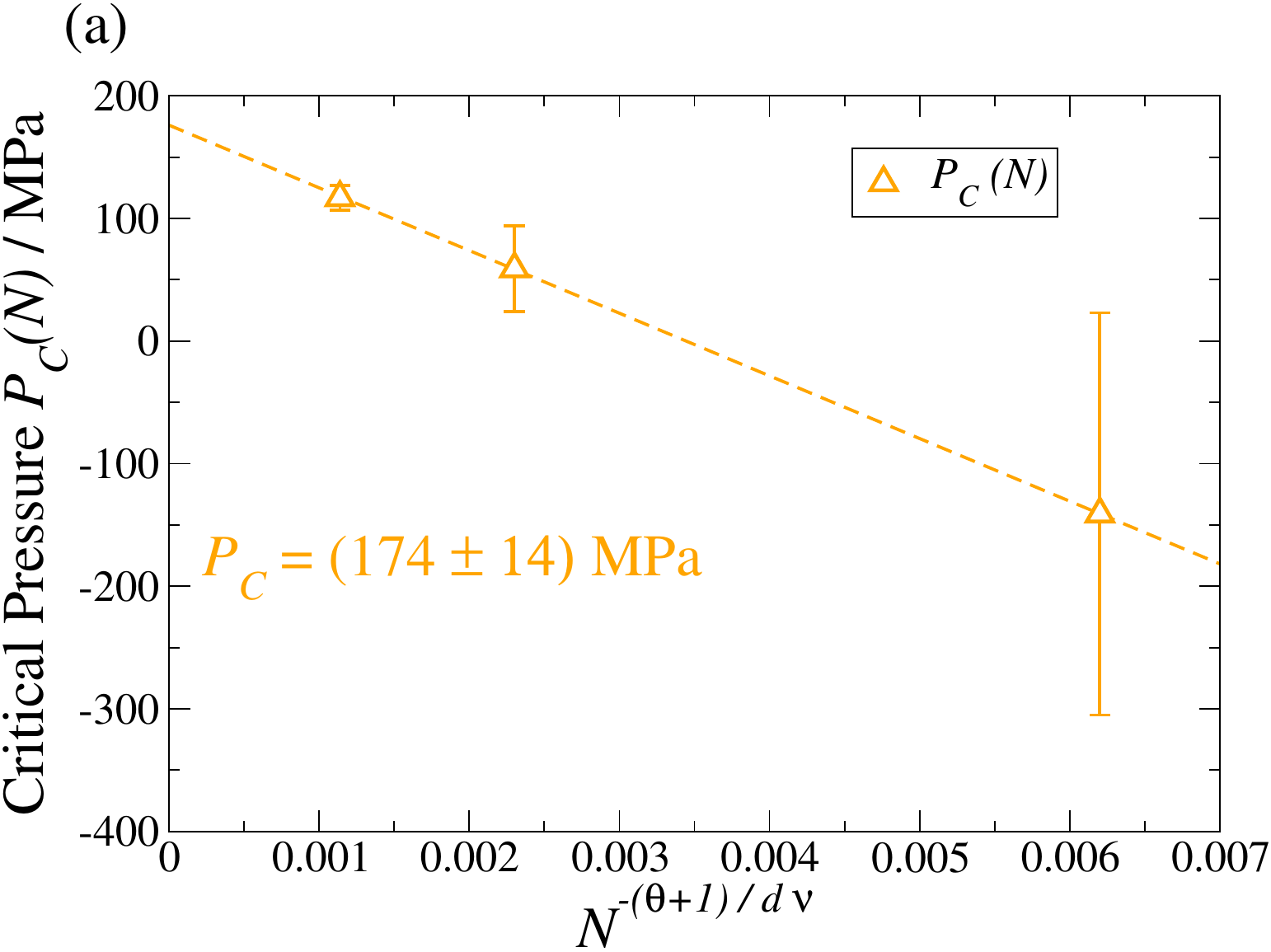} 
\includegraphics[scale=0.34]{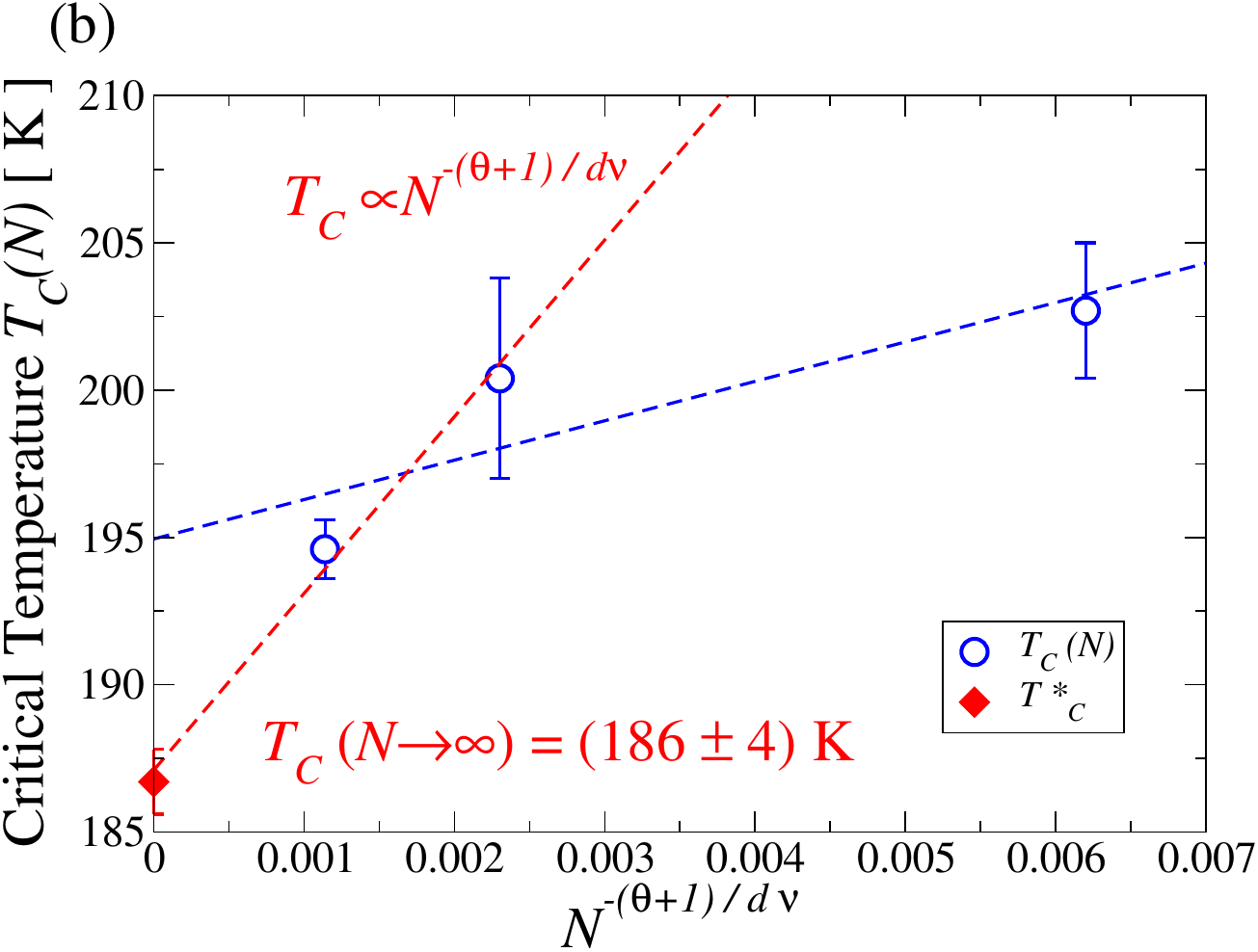}
\caption{{\bf The scaling laws for the critical parameters are consistent with the 3D Ising critical exponents.}
{\bf (a)} $P_C(N)$, in MPa, follows the scaling law in Eq.(\ref{eq:scaling_Tc_Pc}) with 3D Ising critical exponents for the three sizes considered here,  with ${\cal A}_P= (-5 \pm 3)10^5$ MPa and
  $P_C = (176 \pm 14)$ MPa.
{\bf (b)}  $T_C(N)$ (blue circles), in K, deviates from the expected scaling law. However, by thermodynamic argument (see text), we can add the critical temperature $T_C(P_C^\infty)|_{\rm extrema}=(186.7\pm1.1)$~K (red diamond) corresponding to $C_P^{\rm Max}(P)$ at $P_C(N\to\infty)$. Excluding $T_C(N=512)$ and adding $T_C(P_C^\infty)|_{\rm extrema}$, the calculated critical temperatures follow the 3D Ising critical scaling law in Eq.(\ref{eq:scaling_Tc_Pc}) with ${\cal A}_T= (6 \pm 1)10^3$ K and
  $T_C=(186\pm4)$ K (red dashed line).}
    \label{fig:extrapolation_Tc_Pc}
\end{figure}

\begin{table}
\caption{\label{table:critical_parameters} Parameters adopted in Fig.~\ref{fig:Ising3D}. For each $N$, we adjust the critical parameters $T_C$, $P_C$, and $s$,  as in section \ref{sec:estimationLLCP}, to best fit the 3D Ising distribution $Q_3$.}
\begin{ruledtabular}
\begin{tabular}{cccc}
N&$T_C$ [K]& $P_C$ [MPa] &$s$\\
\hline
512  & $202.7 \pm 2.3$ & $-141 \pm 164$      & $-36\pm 30$\\
1,728  & $200.4 \pm 3.4$ & $59 \pm 35$    & $-76\pm 38$\\
4,096 & $194.6 \pm 1.0$  & $117 \pm 10$  & $-142 \pm 15$\\
\end{tabular}
\end{ruledtabular}
\end{table}

We analyze how $T_C$ and $P_C$ extrapolate to the thermodynamic limit. This analysis is crucial to determine whether the observed  LLCP results from finite-size effects or is an intrinsic property of the model. According to the finite size scaling theory~\cite{Wilding1996}, the scaling laws of $T_C(N)$ and $P_C(N)$ are governed by the critical exponents of the universality class as
\begin{equation}
\label{eq:scaling_Tc_Pc}
\begin{split}
  P_C(N) &= {\cal A}_P N^{-(\theta+1)/d\nu} + P_C\\ 
  T_C(N) &= {\cal A}_T N^{-(\theta+1)/d\nu} + T_C
\end{split}
\end{equation}
where ${\cal A}_P$ and ${\cal A}_T$ are scaling constants, $d$ is the dimensionality, $\nu$ the critical exponent controlling the divergence of the correlation length $\xi$, $\theta$ the correction to scaling, and $P_C\equiv P_C(N\to\infty)$ and $T_C\equiv T_C(N\to\infty)$ are the critical pressure and temperature, respectively, in the thermodynamic limit. For the 3D Ising model, it is $d=3$, $\nu=0.63$, and 
$\theta=0.53$ ~\cite{Liu2010}.

By fitting the scaling laws of $P_C(N)$ and $T_C(N)$ to the three calculated LLCP$(N)$, we find that $P_C(N)$ follows the expected power law (Fig.~\ref{fig:extrapolation_Tc_Pc}a), but $T_C(N)$ apparently does not (Fig.~\ref{fig:extrapolation_Tc_Pc}b).
However, the correct dependence of $P_C(N)$, extrapolating to $P_C=(174\pm14)$~MPa, suggests that $T_C(N)$ has a stronger $N$-dependence than the critical pressure, inducing a significant deviation at the smallest size $N=$512. 

To account for this, we observe that the LLCP in the thermodynamic limit must occur at a temperature $T_C(P_C^\infty)|_{\rm extrema}$ along the locus of extrema of the response functions at the extrapolated $P_C(N\to\infty)$. From Fig.s \ref{fig:Supercooled_CP}, \ref{fig:Supercooled_AlphaP}, and \ref{fig:Supercooled_KT}, we estimate $T_C(P_C^\infty)|_{\rm extrema} = 186.7\pm1.1$~K. By adding $T_C(P_C^\infty)|_{\rm extrema}$ and neglecting $T_C(N=512)$, we find agreement with the expected scaling law in Eq.(\ref{eq:scaling_Tc_Pc}) with $T_C=(186\pm4)$~K (Fig.~\ref{fig:extrapolation_Tc_Pc}b).
Here, the error is likely underestimated for the added point and is taken as the largest among the data used in our fit. 
Therefore, we confirm that the water LLCP has the 3D Ising critical exponents and conclude that its critical parameters in the thermodynamic limit are $P_C=(174\pm14)$~MPa, and $T_C=(186\pm4)$~K.

\section{\label{sec:Discussion}Discussion}

Remarkably, our calculations are consistent with a recent overall analysis of available experimental data concluding that the LLCP, if present, should be located between 180 and 200 MPa and at a temperature close to 190 K \cite{Mallamace2024}. These critical parameters are compatible with our estimates, especially considering the approximations adopted in Ref.~\cite{Mallamace2024} and our extrapolation of $T_C$. 

In particular, by recompiling data for the NMR proton chemical shift,
the authors of Ref.~\cite{Mallamace2024} indicate a structural transition along a locus in the 205 K - 220 K temperature range below 200 MPa that disappears at higher $P$. This locus of maxima is a proxy for the Widom line because the variation of NMR proton chemical shift with $T$ along isobars is proportional to the isobaric $T$-derivative of $N_{\rm HB}$ \cite{Mallamace2008} and, as discussed above, the latter is contributing to the extrema of the response functions.
Therefore, our results, showing response functions extrema between 205 K and 220 K, are quantitatively consistent, within the statistical errors, with the experimental data in Ref.~\cite{Mallamace2024} over the entire range of positive pressures we explored. 

It is also intriguing to note that experiments conducted on micro-sized water droplets \cite{Kim978} reveal structural changes in liquid water at temperature $(205\pm10)$ K. These structural changes are interpreted as consistent with crossing an LLPT at a pressure between 1 atm and 350 MPa. The calculations presented here show that, at that temperature and up to approximately 50 MPa, water exhibits the largest structural change marked by the locus of $C_P^{\rm Max}$, which is a proxy to the Widom line \cite{FS2007}. Furthermore, along this line, we find that finite systems undergo macroscopic fluctuations extending over more than 10 nm, the approximate size of our largest sample.

\begin{figure}[!ht]
    \includegraphics[width=0.45\textwidth]{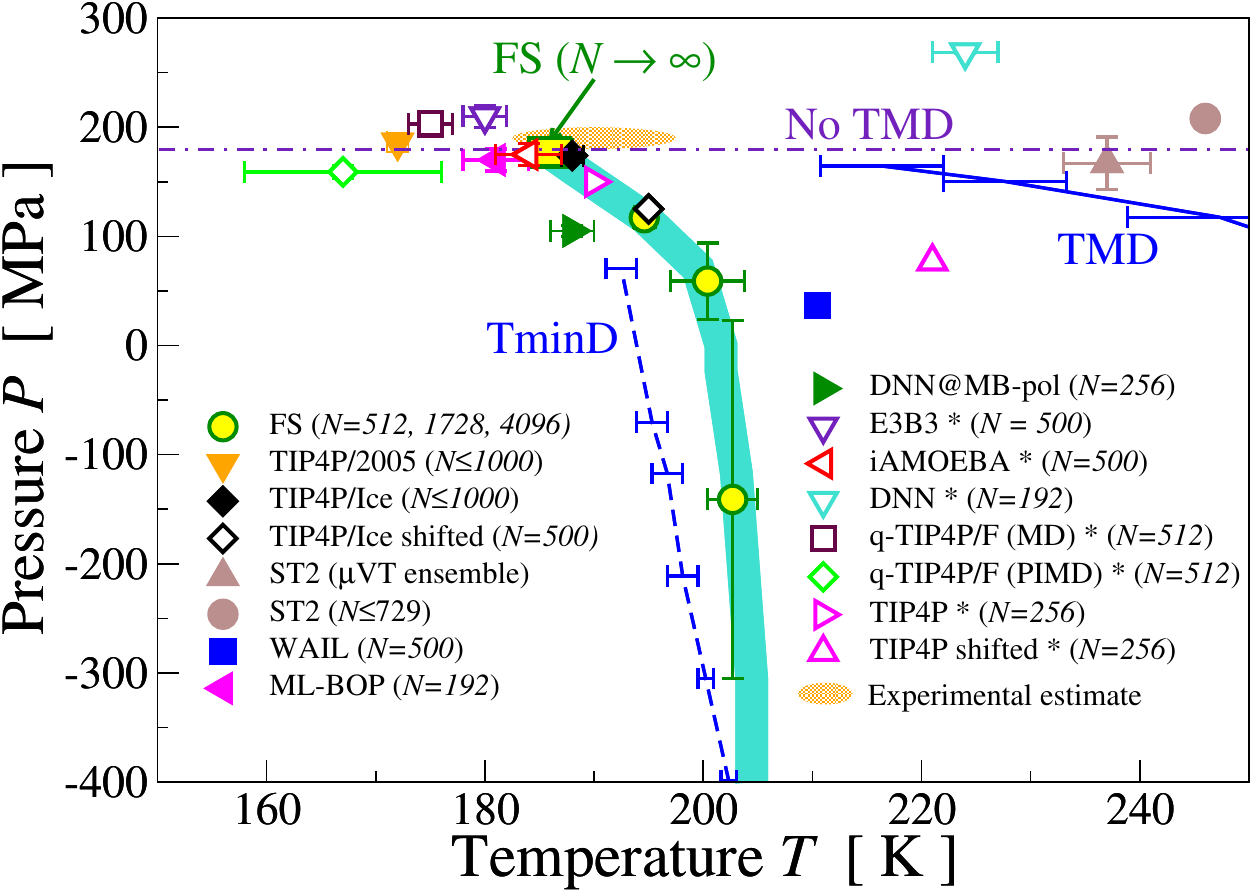}
\caption{{\bf The estimate of the LLCP in the thermodynamic limit compared to other models' predictions.}
Our estimates of the critical parameters for different sizes $N$ (yellow circles, from bottom $N=$ 512 to top 4,096) extrapolate in the thermodynamic limit to $T_C=(186\pm4)~$K and $P_C=(174\pm14)$~MPa (yellow rectangle) close to the pressure $P_{\rm TMD}^{\rm Max}\simeq 180$~MPa above which liquid water, both in the experiments \cite{Mishima:2010fk, Mallamace2024} and in our calculations (Fig. \ref{fig:density_TMD_limit}), displays no TMD (purple dot-dashed line).
The finite-size estimates fall along the locus of $C_P^{\rm Max}(P)$ from Fig.~\ref{fig:Supercooled_CP} (thick turquoise line).
Other estimates of the LLCP are for the
TIP4P/2005 and TIP4P/Ice~\cite{Debenedetti289}, 
TIP4P/Ice shifted~\cite{Espinosa2023},
ST2 ($\mu$VT)~\cite{Liu2009}, 
ST2 ($N\leq 729$)~\cite{Kesselring2013},
WAIL~\cite{Weis:2022ug}, 
ML-BOP~\cite{Dhabal_Kumar_Molinero_2024},
DNN@MB-pol~\cite{Sciortino_Zhai_Bore_Paesani_2024},
E3B3~\cite{Ni2016},
iAMOEBA~\cite{10.1063/1.4963913},
DNN~\cite{Gartner2020},
q-TIP4P/F (MD and PIMD)~\cite{Eltareb:2022wk}, 
TIP4P and TIP4P shifted\cite{corradini:134508},
with sizes and symbols indicated in the legend.
Asterisks ($^*$) in the legend indicate approximate calculations.
The orange-shaded area corresponds to the approximate estimate of the LLCP based on a recent analysis of a collection of experimental data in Ref.~\cite{Mallamace2024}.}
    \label{fig:LLCP_otherModels}
\end{figure}

When compared with other numerical models, our LLCP prediction
 is in good agreement with the approximate estimates from a few others, e.g., the classical polarizable iAMOEBA \cite{10.1063/1.4963913}  (Fig.~\ref{fig:LLCP_otherModels}). 
The iAMOEBA is parameterized using experimental data and high-level {\it ab initio} calculations, where cooperative effects are included \cite{Wang_amoeba_2013}. The analysis of $K_T$ and $C_P$ from atomistic simulations of the iAMOEBA suggests that it exhibits an apparent divergent point around 
$(175\pm10)$~MPa, and $(184\pm3)$~K \cite{10.1063/1.4963913}.

More recently, free energy calculations for 192 molecules of a machine-learned monatomic water model with three-body interactions (ML-BOP) for both liquid water and ice \cite{Chan2019} have shown the occurrence of the LLCP at 
$P_C=(170\pm10)$~MPa and $T_C=(181\pm3)$~K \cite{Dhabal_Kumar_Molinero_2024}. 
This estimate is very close to our result.

Notably, both the iAMOEBA and the ML-BOP models include optimized many-body interactions (HB cooperativity).
This is true also for our FS model whose parameters were selected \cite{Coronas-2024} based on {\it ab initio} energy decomposition analysis of small water clusters \cite{KhaliullinKuhne2013}.
Therefore, the agreement of our LLCP estimate with those from other cooperative models might be because they have a similar intensity of HB cooperativity, consistent with previous research indicating the importance of many-body interactions in determining the LLCP \cite{Stokely2010}. 

On the other hand, our prediction of the LLCP is within the statistical error of the rigid TIP4P/Ice model, whose o. p. fluctuations analysis leads to an estimate of $P_C=(173.9\pm0.6)$~MPa and $T_C=(188\pm1)$~K \cite{Debenedetti289}.
The TIP4P/Ice model considered in Ref.~\cite{Debenedetti289} is parametrized to reproduce the crystal phase diagram of water \cite{Abascal2005}. 
However, Espinosa et al. have shown that the model must be shifted by 40 MPa to fit the liquid water equation of state and compressibility, leading to a different estimate of the LLCP around 125 MPa and 195 K \cite{Espinosa2023}.
This new prediction falls along the locus of $C_P^{\rm Max}(P)$ as our finite-size LLCP estimate for $N=$4,096 FS water molecules. 
The absence of finite-size scaling analysis in Ref.~\cite{Espinosa2023} prevents a consistent comparison with the original TIP4P/Ice model \cite{Debenedetti289} and the present work.

Similar considerations hold also for the TIP4P model, for which the LLCP has been estimated around 190 K and 150 MPa \cite{corradini:134508}, a state point which falls along our $C_P^{\rm Max}(P)$ line. However, when the model is shifted \cite{corradini:134508} by $+31$ K and $-73$ MPa to match the TMD experimental data, the LLCP estimate moves far from our calculations.

Finally, we observe that the average value of our prediction for the critical pressure $P_C=(174\pm14)$~MPa falls below the limiting pressure for the density anomaly, $P_{\rm TMD}^{\rm Max}\simeq 180$~MPa, calculated in the framework of our model (Fig. \ref{fig:density_TMD_limit}). Experiments show a consistent behavior, with the TMD no longer evident above 180 MPa \cite{Mishima:2010fk, Mallamace2024}. 
Additionally, there is a drastic change in the slope of the melting line from negative to positive in the $(T, P)$ plane, below and above 200 MPa, respectively, as well as for the temperature of the homogeneous crystallization \cite{Mallamace2024}. 

The negative slope is due to the density anomaly. It is a result of the anticorrelation between volume variation $\Delta V$ and entropy change $\Delta S$ at the melting line for the Clausius-Clapeyron equation, which is given by $dP/dT|_{\rm phase~transition}=\Delta S/\Delta V$. A positive slope of the melting line is typical of normal liquids without density anomaly. 

We expect a negative slope also for the LLPT as there is a similar $\Delta S/\Delta V$ between the high-$T$ HDL and the low-$T$ LDL phase. Therefore, in the thermodynamic limit, the LLPT should occur below the pressure where the melting line of water changes its slope and there is no TMD, between 200 MPa and $P_C=(174\pm14)$~MPa (Fig. \ref{fig:LLCP_otherModels}). 

It's fascinating to see that the TMD and TminD extrapolations are consistently converging toward our estimate of the LLCP (Fig. \ref{fig:LLCP_otherModels}). This convergence suggests that the density anomaly region, which is delimited by the TMD and TminD lines, might originate from the LLCP, along with the Widom line. If this prediction is confirmed, it would provide compelling evidence of the relationship between the LLCP and the anomalies of water.

\section{\label{sec:Conclusions}Summary and conclusions}

Experimental studies of the metastable region of liquid water are challenging \cite{Kim978, doi:10.1073/pnas.2018379118, Holten:2017um} but fundamental for discriminating between different thermodynamic explanations of the water properties.
Therefore, it is crucial to investigate the extrapolation of the equation of state of reliable models to the experimentally unexplored regions to address open questions related to the possible different thermodynamic scenarios \cite{llcp, Stokely2010, Debenedetti289}.

In this work, we consider the FS liquid water. This model provides a precise quantitative reproduction of the experimental data for the density, specific heat $C_P$, thermal expansion coefficient $\alpha_P$, and compressibility $K_T$ of liquid water with an unprecedented level of accuracy around ambient conditions over a temperature range of roughly 60 degrees at atmospheric pressure and up to 50 MPa \cite{Coronas-2024}. As a result, it is one of the most reliable molecular models for liquid water at equilibrium. Here, we examine its phase diagram upon supercooling and stretching.

We analyze how the response functions, $C_P$, $\alpha_P$, and $K_T$, change at supercooled temperatures by varying the pressure from negative to positive. We find the loci of response functions extrema and verify their consistency with general thermodynamic relations and the available experimental data. 

Our results show that all the loci of extrema converge, within the statistical errors, to one single line that extends from positive pressure to moderately negative for the finite-size systems we consider here. Along the common locus of the extrema, we find that the liquid flips in energy and density between HDL-like and LDL-like states, as expected along the LLPT line. Furthermore, the Gibbs free energy calculated at the end of this line displays two minima as near an LLCP between the LDL and HDL.

Through a detailed scaling analysis, we find that the finite-size effects are significant. Still, the water LLCP also exists in the thermodynamic limit and belongs to the 3D Ising universality class. The critical parameters extrapolate to $T_C=(186\pm4)$~K and $P_C=(174\pm14)$~MPa.

Several recent studies of water models with different degrees of coarse-graining support the prediction of an LLCP. Our estimate of $(T_C, P_C)$ is consistent with those for two cooperative models that optimize the many-body interactions, the iAMOEBA \cite{10.1063/1.4963913} and the ML-BOP \cite{Dhabal_Kumar_Molinero_2024}. General arguments \cite{Stokely2010} suggest that they have an intensity of HB cooperativity comparable to that of the FS model. Intriguingly, these three models are based on a combination of experimental data and {\it ab initio} or machine-learning optimizations. They stand out for their quantitative agreement for the water density, although the FS outperforms the others for the response functions. Therefore, the FS model is reliable around ambient conditions and also comparable to optimized models in the metastable region.

The consistency among these models that optimize the many-body interactions and agree quantitatively with the water equation of state highlights the importance of hydrogen-bond cooperativity in explaining water anomalies. Previous theoretical studies have shown that scenarios without the LLCP are thermodynamically consistent and can also explain the water anomalies. However, these scenarios are excluded when cooperativity is considered and has a moderate intensity as in the present model \cite{Stokely2010}.

Our LLCP estimate is also consistent with that of the TIP4P/Ice optimized for the solid phases \cite{Debenedetti289}, but not with the TIP4P/Ice shifted, optimized for liquid water \cite{Espinosa2023}. However, the latter has critical parameters similar to our calculations for finite systems. 

Furthermore, experiments \cite{Kim978} on supercooled water droplets show significant structural changes at pressures and temperatures corresponding to our estimate for the $C_P$ maxima. Therefore, our work suggests an explanation for the results in Ref. \cite{Kim978} that, although alternative, is still consistent with the LLCP's existence.

However, what is even more compelling is that our prediction for the LLCP matches, within the statistical error, with the estimate based on a comprehensive set of measurements recently analyzed \cite{Mallamace2024}. The study also reviews data demonstrating significant ordering of the HB network over a range of pressures and temperatures \cite{Mallamace2024}, which are consistent with our estimates for the HB structuring, the extrema of the response functions, and the Widom line.

Finally, our prediction for the LLCP approaches the pressure limit where no TMD is found in the model \cite{Coronas-2024} and the experiments \cite{Mishima:2010fk, Mallamace2024} and the melting line changes its slope. Based on this, the LLPT should occur between 200 MPa and $P_C=(174\pm14)$~MPa and it should lie with the LLCP at the convergence of the TMD and the TminD lines, providing persuasive evidence on the origin of the anomalies of water.

On a more applicative side, this study shows that a water model with a) high quantitative accuracy and transferability in a broad thermodynamic range around ambient conditions, b) capability of large-scale free-energy calculations, and c) computationally efficiency, also is able to explore the metastable phases of water providing reliable calculations.

Our findings indicate that the hydrogen bond network can develop significant cooperative fluctuations at scales larger than 10nm when exposed to supercooled temperature of around 205 K and pressure up to 50 MPa. These thermodynamic values are widely used in biostorage technology for cryopreservation of genetic material, biological tissues, and medications. Hence, it is crucial to understand the density fluctuations under these conditions to ensure long-term biopreservation without causing any cryo-injury \cite{GUO202453}.

Furthermore, an advantage of the FS model is its ability to efficiently calculate free energy for systems containing millions of molecules. This feature is particularly useful for the investigation of large hydrated systems in various fields without requiring expensive computational resources or long real-time simulations. Such studies can address important questions relevant to nanotechnology \cite{Kavokine:2022ue, BELLIDOPERALTA2023123356, SPINOZZI2024446} and biophysics \cite{FranzeseSCKMCS2008, Camilloni2018Advanced}, which often face limitations with conventional computational approaches \cite{Best:2014wg, Piana:2015aa, Huang:2017aa, Aydin2019}. While the FS model is currently optimized only for bulk water, it has demonstrated great potential in these fields, providing many qualitative and semi-quantitative results \cite{Bianco:2015aa, BiancoPRX2017, Bianco:2017ab, Bianco-Navarro2019, Bianco:2020aa, polym13010156, Fauli:2023aa}. Further studies are required to achieve quantitative predictions with hydrated interfaces.

\begin{acknowledgments}
L.E.C. acknowledges support from the Universitat de Barcelona grant no. 5757200 APIF\_18\_19. 
G.F. acknowledges the support by MCIN/AEI/ 10.13039/ 501100011033 and “ERDF A way of making Europe” grant number PID2021-124297NB-C31, by the Ministry of Universities 2023-2024 Mobility Subprogram within the Talent and its Employability Promotion State Program (PEICTI 2021-2023), and by 
 the Visitor Program of the Max Planck Institute for The Physics of Complex Systems for supporting a visit started in November 2022.
\end{acknowledgments}

\section*{Author declaration}

\subsection*{Conflict of interest}

The authors have no conflict of interest to disclose.

\subsection*{Author contributions}
L.E.C.: Data curation (equal), Formal analysis (equal), Investigation (equal), Methodology (equal), Software (lead), Writing – original draft (equal), Writing – review \& editing (equal).
G.F.: Conceptualization (lead), Data curation (equal), Formal analysis (equal), Funding acquisition (lead), Investigation (equal), Methodology (equal), Project administration (lead), Resources (lead), Software (supporting), Supervision (lead), Writing – original draft (equal), Writing – review \& editing (equal).

\section*{Data Availability Statement}

The code and analysis script to reproduce the findings of this study are openly available at https://github.com/lcoronas/FSBulkWater/.
The data that support the findings of this study are available from the corresponding author upon reasonable request.

\appendix

\section{\label{app:TheFSModel}The FS model}

A detailed description of the model with a full justification for its parameters is given in Ref. \cite{Coronas-2024}. Here, we briefly define the model, its variables, and its parameters.

The model assumes that, at constant temperature $T$ and pressure $P$,  $N$ water molecules are distributed in a variable volume $V(T, P)$. The entire $V$ is partitioned into cells $i \in [1, ... ,{\cal N}]$,
with ${\cal N} \geq N$, each containing at most one molecule. 
Here, we set ${\cal N} = N$.
Each cell has volume $v \equiv V/{\cal N} \geq v_0$,
where $v_0 \equiv r_0^d$ is the van der Waals volume of one molecule, with $r_0\equiv 2.9$~\AA\  and $d$ is the Euclidean dimension.

The FS model coarse-grains the positions of the molecules,
with $r \equiv v^{1/d}$  corresponding to the average distance between two neighboring molecules.
The van der Waals interaction, modeled with the Lennard-Jones potential with an energy parameter $\epsilon \equiv 5.5 kJ/mol$ \cite{Coronas-2024}, determines the value of $v(T, P)$ and the homogeneous component of the density $v_0/v$, that is discretized with an index $n_i  = 0$ if $v_0/v \leq 0.5$ (gas-like density), and $n_i  = 1$ (liquid-like density) otherwise, for each cell and molecule $i$. The heterogeneous component of the density is associated with the number of HBs each molecule forms, as we discuss in the following.

The model separates the HB interaction into two Hamiltonian terms: the directional (covalent) component,
and the cooperativity component, much weaker than the first and it is due to many-body correlations. They are described in terms of a
set of {\it bonding}
$\sigma_{ij}$ variables.
Each water molecule $i$ has a bonding variable
$\sigma_{ij} \in [1,...,q]$ for each water molecule $j$ that is its
 neighbor. 
 Consistent with the experiments,
 the model assumes that each water molecule can form only four tetrahedral HBs. 
 
 A HB is formed only if two neighboring molecules have the facing bonding variables in the same state, i.e.,  $\sigma_{ij} = \sigma_{ji}$. 
 Debye-Waller factors estimates \cite{Teixeira1990} and calculations \cite{Ceriotti:2013aa} show that 
only $1/6$ of the entire range of possible values $[0,360^\circ]$ of the ${\widehat{\rm OOH}}$  angle between two molecules is associated with a bonded state. Therefore, the model sets $q=6$ to account for this constrain. 

Furthermore, consistent with calculations  \cite{Luzar-Chandler96, Hus:2012uq, Schran2019}, the HB has a negligible energy, i.e., is broken, for O--O larger than a distance $r_{\rm max}$. The FS model sets $r_{\rm max}\simeq 3.65$\AA. This choice guarantees that, in the homogeneous system, the HB breaks if the O--O distance between two bonded molecules becomes $r>r_{\rm max}$, i.e., $v_0/v\equiv r_0^3/r^3<0.5$, then $n_i = 0$ $\forall i$.  If, instead, $r\leq r_{\rm max}$ then $n_i = 1$ $\forall i$, and $n_i n_j =1$, allowing the HB formation. Hence, $n_i$ is a {\it bonding} index.

In a cubic lattice, each molecule has six nearest neighbors. However, a water molecule cannot form more than four HBs. For this reason, we
introduce a set of {\it allowing} variables $\eta_{ij}\in \{0, 1\}$, where $0$ denotes a forbidden bond between molecules $i$ and $j$, and 1 denotes an allowed bond. By construction, for each molecule $i$, four of the
six variables $\eta_{ij}$ are set to 1, while the other two are 0, with
$\eta_{ij} = \eta_{ji}$ $\forall ij$. Ref.~\cite{CoronasThesis} describes an algorithm for generating valid configurations of the $\eta_{ij}$ variables. 

The total number of HBs is, therefore,
\begin{equation}\label{N_HB}
N_{\rm HB} = \sum_{<i,j>} n_i n_j \eta_{ij} \delta_{\sigma_{ij}, \sigma_{ji}}
\end{equation}
where the sum is over the nearest neighbor molecules and $\delta_{a,b} = 1$ if $a = b$, and $0$ otherwise.
The Hamiltonian term corresponding to the directional component of the HB interaction is pair-additive and linear in $N_{\rm HB}$,
\begin{equation}\label{eq:Ham_HB}
  {\cal H}_{\rm HB} \equiv -J N_{\rm HB},
\end{equation}
where $J$ is set to 11 kJ/mol \cite{Coronas-2024}.

The cooperative component of the HB interaction is given by a five-body term, consistent with the analysis of polarizable water clusters \cite{Abella:2023ab},
\begin{equation}\label{eq:Ham_sigma}
  {\cal H}_{\sigma} \equiv -J_\sigma N_\sigma \equiv -J_\sigma \sum_i n_i \sum_{(k,l)_i} \delta_{\sigma_{ik}, \sigma_{il}},
\end{equation}
where $(k,l)_i$ indicates all possible pairs of bonding variables of the $i$-th water molecules, and $J_\sigma=1.76$ kJ/mol is set\cite{Coronas-2024} based on ALMOEDA analysis \cite{KhaliullinKuhne2013} of density functional theory calculations on water clusters \cite{C2CP42522J}. 

Finally,  the total volume of the system is 
\begin{equation}
\label{eq:Vtot}
V_{\rm Tot} \equiv V +v_{\rm HB} N_{\rm HB}, 
\end{equation}
where $v_{\rm HB}=14.4$~\AA$^3$ accounts for the local decrease in density due to HB formation \cite{Skarmoutsos:2022ur}. Hence, $V_{\rm Tot}$ includes the local heterogeneity in the density field due to the HBs.

\section{\label{app:OrderParameter}The order parameter}

We consider $Q(\bar{e},\bar{\rho})$, where $\bar{e}$ and $\bar{\rho}$ are dimensionless quantity, and rotate it using the 2D Euclidean rotation matrix,  ${\cal M}=\bar{e}\sin(\alpha)+\bar{\rho}\cos(\alpha)$. 
Therefore, we define
$M (s) \equiv {\cal M}/\cos(\alpha) = \bar{\rho} + s \bar{e}$,  
where $s \equiv \tan(\alpha)$. 

Our joint probability distribution $Q(e,\rho)$ has a strongly asymmetric shape, as evident from the asymmetry between the two basins of attractions of $\Delta G$ (Fig.\ref{fig:HR}), considering that $Q(e,\rho)=\exp[-\Delta G(e,\rho)/k_BT]$.
Specifically, the basin for the LDL state is broader than that for the HDL.

Therefore, to make the two basins more symmetric, we need to consider 
only rotations for $\alpha\in [\pi/2, \pi]$ and $[3\pi/2, 2\pi]$, hence 
$s<0$. We check that the sign of $s$ depends on the model's parameter choice.

\begin{figure}
    \centering\includegraphics[scale=0.24]{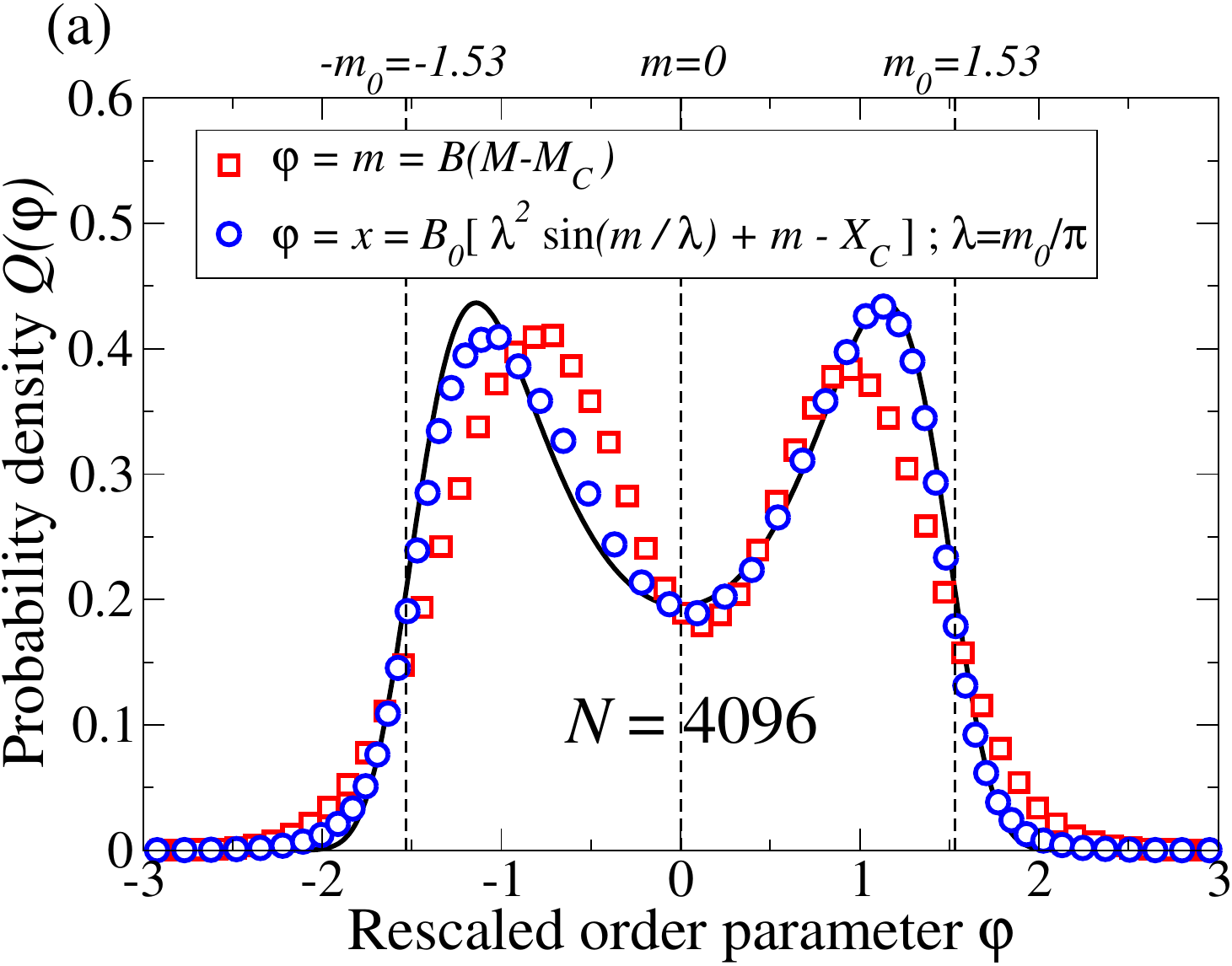}
    \includegraphics[scale=0.24]{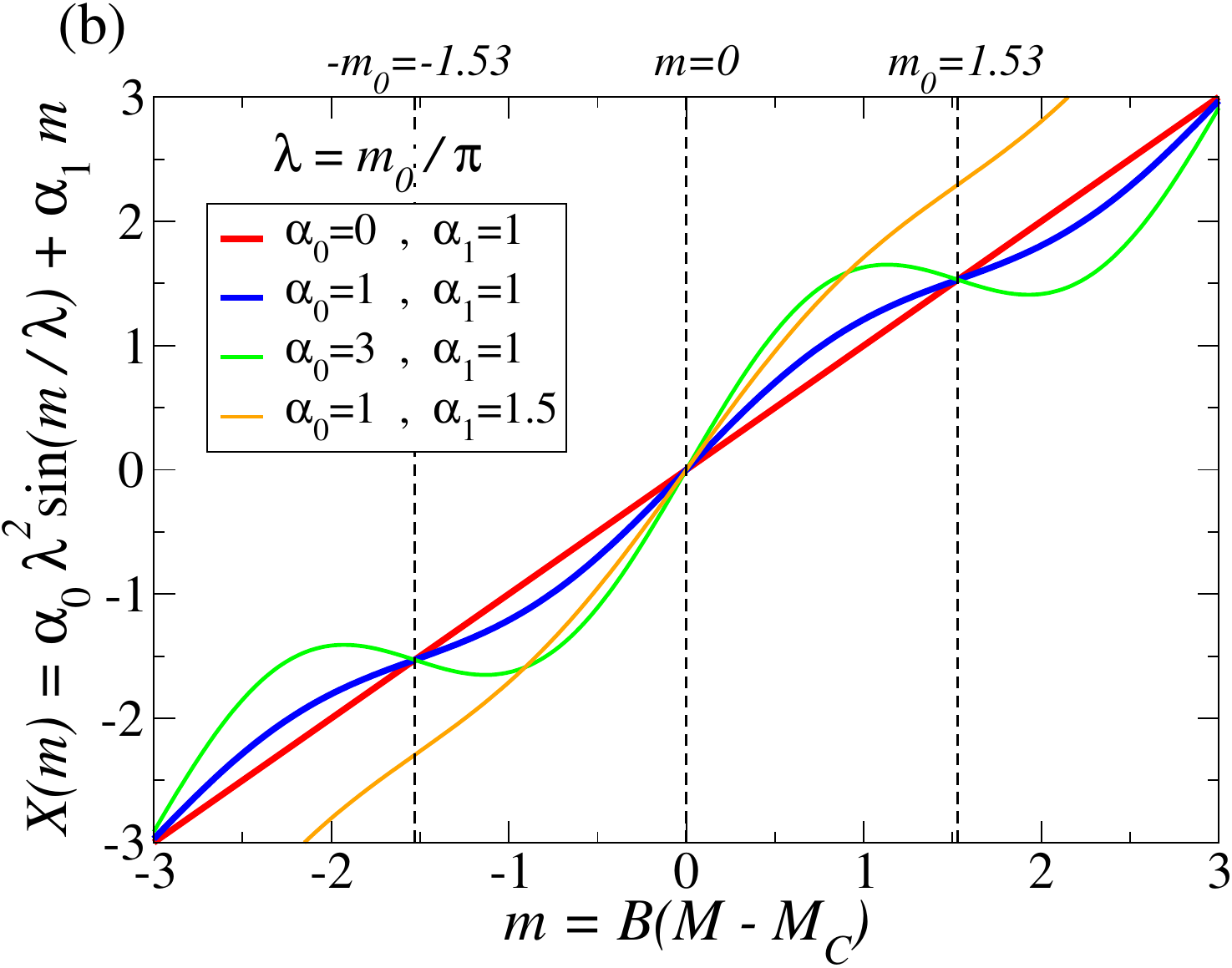}
\caption{{\bf Rescaling of the order parameter $M$ and comparison with $Q_3$.}
(a) Comparison between the probability distribution $Q(m)$ (red squares) for the rescaled $m\equiv B(M-M_C)$;
$Q(x)$ (blue circles) for the rescaled o. p. $x$ defined in Eq.(\ref{x}); and 
$Q_3(\varphi)$ (black line).
Symbols are calculated for the FS model with $N=$4,096 near the estimated $(T_C(N), P_C(N))$. 
$Q(m)$ coincides with $Q_3$ at $m=0$ and $m_0=\pm 1.53$.
(b) The function $X(m)$ in Eq.(\ref{capital_x}) modulates $m$ allowing us to best fit $Q$ to $Q_3$.
All the curves are for $\lambda=m_0/\pi$ (con $m_0=1.53/\pi$ for $N=$4,096) so that, when $\alpha_1=1$, $X(m)=m$ at $m=0$ and $\pm m_0$ (marked by the dashed vertical lines).
For $(\alpha_0, \alpha_1)=(0,1)$, it is $X(m)=m$ (red line).
For other combinations with $\alpha_1=1$, $\alpha_0$ controls the amplitude of the $m$ modulation.
The combination $(\alpha_0, \alpha_1)=(1,1)$ is our best choice (blue line), and it is used to calculate the distribution $Q$ with blue squares in panel (a).
Other combinations $(\alpha_0, \alpha_1)$ are shown for comparison (green and orange lines).}
    \label{fig:norm_x}
\end{figure}

For each size $N$, we test a candidate $(T_C(N), P_C(N))$ by 
rescaling $M$ as $m\equiv B(M-M_C)$, where $B$ and $M_C$ (see Table \ref{table:critical_parameters}) make $Q(m)$ with zero mean and unit variance. 
However, we realize that any combination of $(T_C(N), P_C(N))$ and $s(N)$ allows us to fit  $Q(m)$ to $Q_3$ only partially. Specifically, we find that the tails of $Q(m)$ are overrated and the peaks are underrated (Fig.~\ref{fig:norm_x}a, red squares).

We, therefore, test alternative functional forms $X(m)$ for the rescaled o. p. in such a way to best fit $Q(X)$ to $Q_3$.
First, we observe that $Q(m)-Q_3=0$ for $m=0$ and $m=\pm m_0$ (Fig.~\ref{fig:norm_x}a).
Thus, we consider only transformations with fixed points in $m=0$ and $\pm m_0$.
Furthermore, $X(m)$ must slightly shrink the tails ($|m|>m_0$) and stretch the peaks ($|m|<m_0$) of our $Q(m)$ to fit $Q_3$
(Fig.~\ref{fig:norm_x}a).

Under these considerations, we find that the function
\begin{equation}
   X(m) \equiv \alpha_0 \lambda^2\sin\left(\frac{m}{\lambda}\right) + \alpha_1 m
   \label{capital_x}
\end{equation}
fits well to our purposes because, when $\alpha_1=1$ and $\lambda = m_0/\pi$, it has fixed points  $\forall m=k\cdot m_0$, with  $k\in\mathbb{Z}$.
Moreover, the sinus alternatively changes its sign shifting $m$ as
convenient, while $\alpha_0$ controls the amplitude of the modulation (Fig.~\ref{fig:norm_x}b). 

Next, by observing that $(\alpha_0, \alpha_1)=(1,1)$ is the best combination and that the modulation requires a further normalization to fit $Q_3$, we get the final expression for the rescaled o. p. as
\begin{equation}
x \equiv  B_0(X(m) -X_C)
   \label{x}
\end{equation}
where the constant $B_0$ and $X_C$ are given in the Table \ref{table:critical_parameters}, together with $\lambda$, for all the values of $N$.


\bibliography{ref}


\begin{table*}
\caption{\label{table:simulation_times} 
To analyze the HDL-LDL transition, for each $T$, we use several MC steps that depend on $N$ and $P$. These MC steps increase by one order of magnitude for decreasing $P$ at constant $N$ and by one or two orders by increasing $N$ at constant $P$. Due to time constraints, we limit our analysis to $N =$ 4096 at $P \geq$ 110. 
As a reference, we also indicate the average real-time corresponding to performing, at the given $N$, the indicated number of MC steps of sequential Wolff update (on $\sigma_{ij}$ variables) and parallel Metropolis (on $\eta_{ij}$ variables) using 
a workstation with a 3.5 GHz CPU XenonW-2155  and a GPU NVIDIA RTX 2080Ti, respectively ~\cite{CoronasThesis}. The variables $\sigma_{ij}$ and $\eta_{ij}$ are defined in Appendix A. Note that the real-time calculations scale linearly with $N$ and MC steps. We tested this scaling by benchmarking systems with $N=10^6$ water molecules. The prediction is that the workstation indicated above will need approximately 2 months in real time to generate $10^5$ MC steps for 100 million water molecules at ambient conditions.}

    \centering
    \begin{ruledtabular}
\begin{tabular}{ cccc}
$N$ & Pressure $P/{\rm ~MPa}$ & Multiple HDL-LDL transitions? & MC steps (Real Time)\footnote{Real Time is indicated for a single $T$. For every isobar, we
simulate at least five temperatures near the critical region.}\\
\hline
\multirow{2}{*}{$512$} & $P  \geq -120$ &  \multirow{5}{*}{YES}  &  $10^5$ (1 min) \\
\cline{2-2}\cline{4-4}
   & $P < -120$ &        &  $10^6$ (10 min)\\
\cline{0-1}\cline{4-4}
\multirow{2}{*}{1,728}& $P \geq -20 $ &   &  $10^7$ (2-3 h)\\
\cline{2-2}\cline{4-4}
    & $P < -20$ & & $10^8$ (20-30 h)\\
\cline{0-1}\cline{4-4}
\multirow{2}{*}{4,096} & $P \geq 110$  &  &  \multirow{2}{*}{$10^8$ (2-3 days)} \\
\cline{2-3}
  &  $P < 110$ &  NO  &  \\

\end{tabular}
\end{ruledtabular}
\end{table*}


\begin{table*}
\caption{\label{table:critical_parameters} 
Set of rescaling parameters such that the probability distributions $Q(m)$ and $Q(x)$, with $m\equiv B(M-M_C)$ and $x\equiv B_0(X-X_C)$, have zero mean and unit variance. The parameters bear no units since $M$ and $X$ are dimensionless. We note that $B_0 \simeq 1$ and $X_C \simeq 0$, being $x\simeq X \simeq m$. We fix $\lambda(N)$; then it has no associated error. All quantities are defined in the main text.}

    \centering
    \begin{ruledtabular}
\begin{tabular}{cccccc}
$N$ & $B$ & $M_C$ & $B_0$ & $X_C$ & $\lambda$ \\
\hline
$512$ & $1.8\pm 0.5$ & $-75 \pm 41 $  & $0.985\pm 0.15$  &  $ -0.027 \pm 0.008 $ &  $1.61/\pi$\\

1,728 & $6\pm 2$ &  $-35 \pm 5 $ &  $0.99\pm 0.02$  & $ 0.010 \pm 0.003 $ &  $1.61/\pi$\\

4,096 & $24\pm 10$ & $-28.5 \pm 0.5 $  &  $1.00 \pm 0.02$  & $ 0.000 \pm 0.001 $ &  $1.53/\pi$\\

\end{tabular}
\end{ruledtabular}
\end{table*}

\clearpage

\begin{figure*}
\includegraphics[scale=0.5]{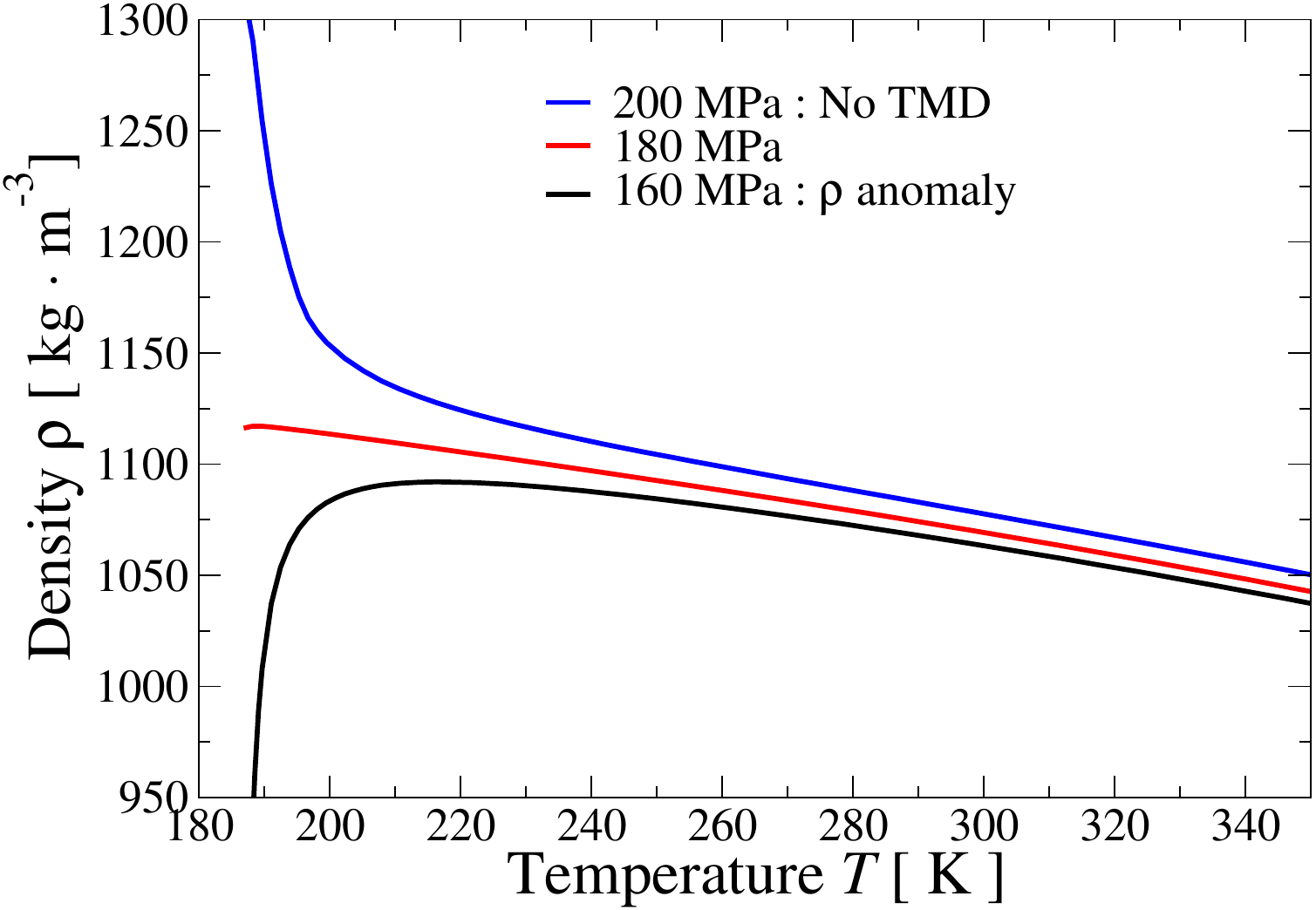}
\caption{{\bf TMD limit.} MC calculations of density for $N =$32,768 water molecules (lines) as in Fig.1a of the main text.
Within our resolution of pressures, the FS water density is anomalous up to 180 MPa. Below 180 MPa, all the isobars display a temperature of maximum density (TMD). At 180 MPa, the TMD approaches the lowest $T$ we explored ($\simeq$ 187 K). At 200 MPa, the isobar monotonically decreases for increasing $T$ as in normal liquids. Within the approximation of the model, given the definition of $V_{\rm Tot}$ in Eq. A4, the choice of the model's parameters and the linear rescaling function defining $P$ in Ref.~\cite{Coronas-2024}, at  $P_{\rm TMD}^{\rm Max}=(J/v_{\rm HB} ~469.459-211.15552)$~MPa$=180.06$~MPa the volume becomes a monotonic function of $T$ along isobars. Therefore, above $P_{\rm TMD}^{\rm Max}$ no TMD can exist within the model. This is quantitatively consistent with the experimental data \cite{Mishima1998, Mallamace2024}.
}
\label{fig:density_TMD_limit}
\end{figure*}

\begin{figure*}
\includegraphics[scale=0.5]{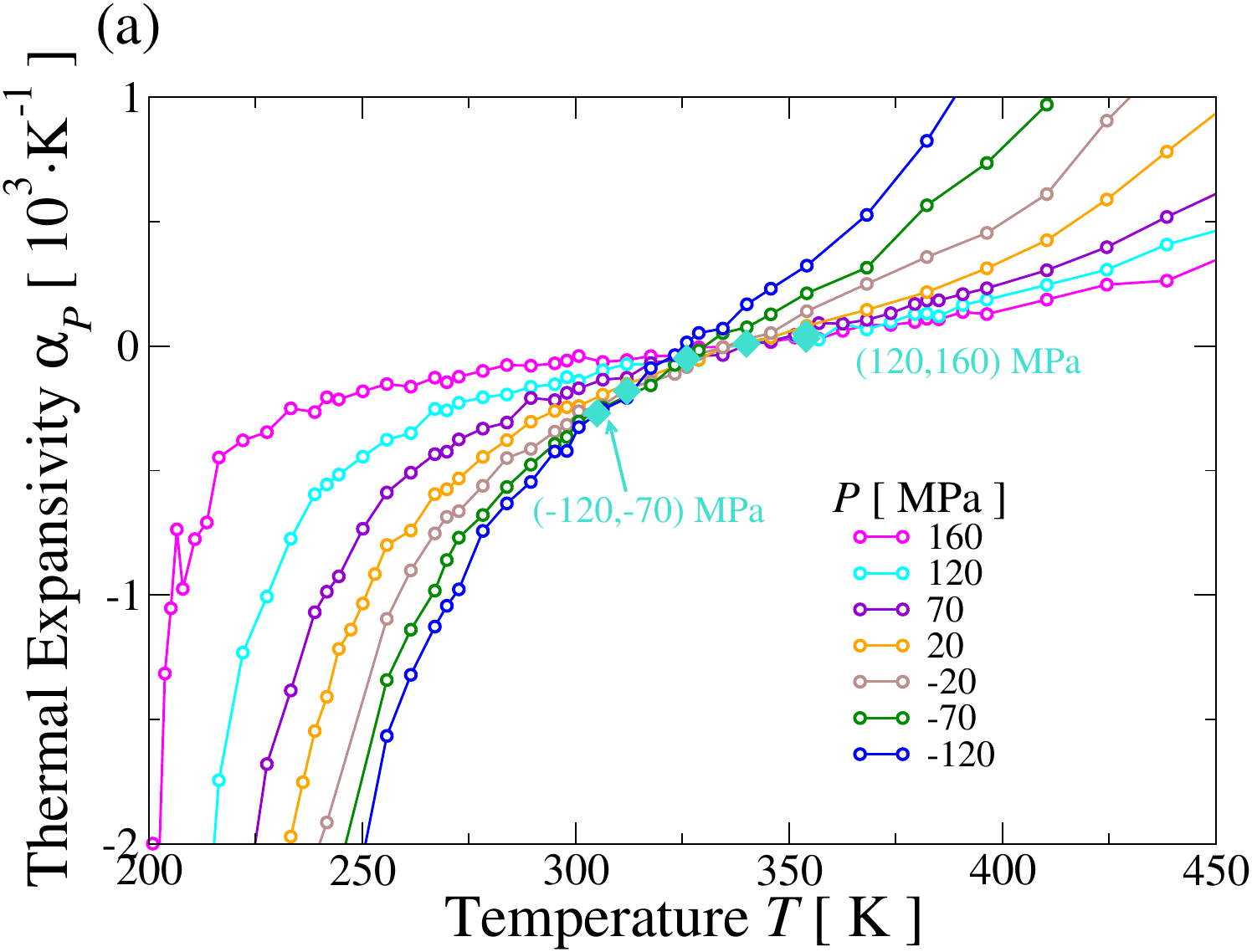}
\includegraphics[scale=0.5]{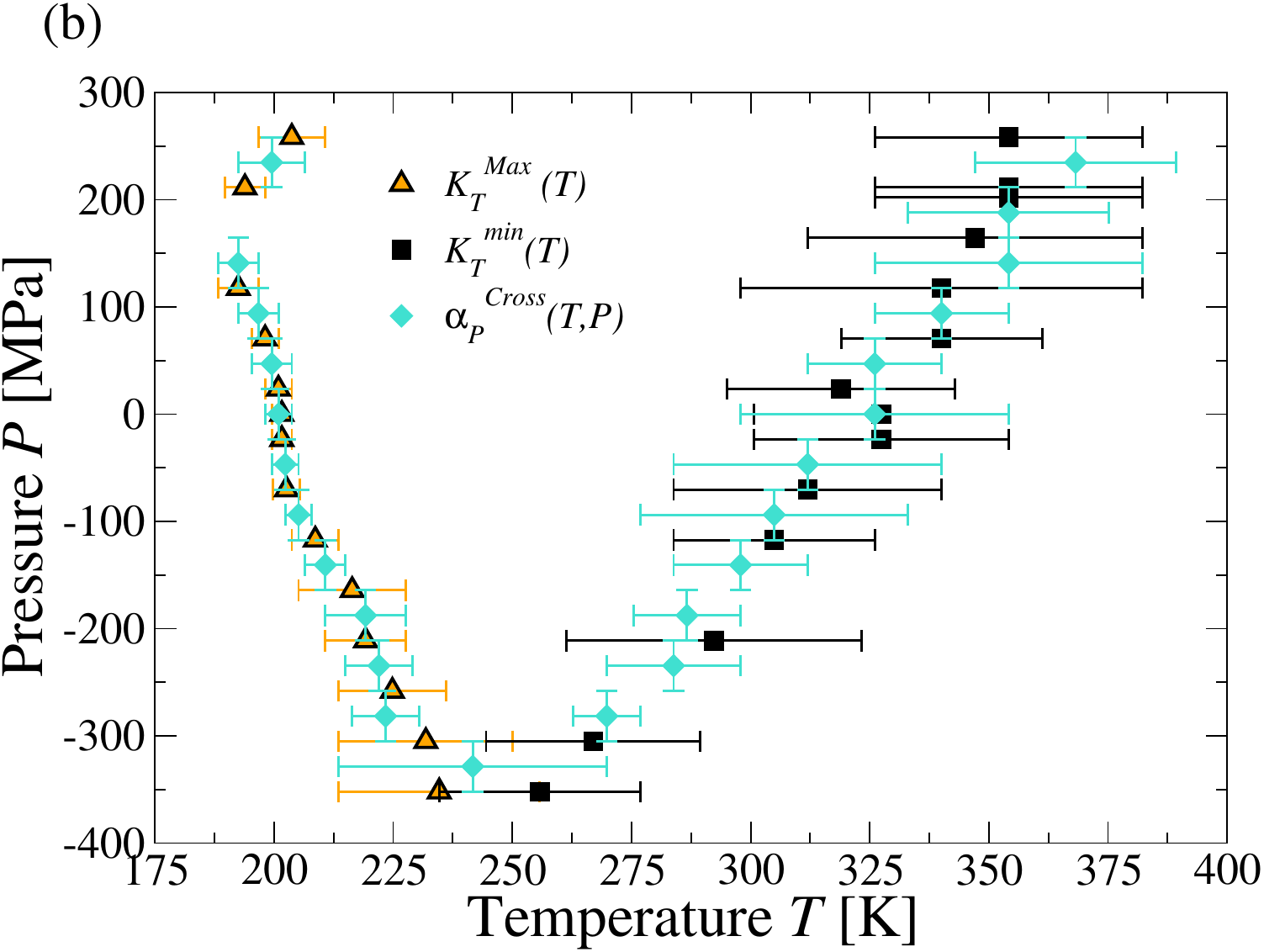}
\caption{{\bf The crossing of $\alpha_P$ and The $K_T$ extrema.} 
{\bf (a)} MC calculations of $\alpha_P$ for $N =$32,768 water molecules (circles), corresponding to Fig.3a of the main text, for pressures between $-120$ MPa and 160 MPa (colors indicated in the legend).
Full turquoise diamonds mark crossing points between consecutive pressures.
{\bf (b)} The same crossing points of panel (a) coincide with the $K_T$ extrema in the $(T, P)$ plane within the error bars, es expected for thermodynamic consistency.}
\label{fig:alpha_crossing}
\end{figure*}

\end{document}